\documentclass[iop, apj]{emulateapj}
\usepackage{graphicx}
\begin{document}
\title{The GALEX Arecibo SDSS Survey V: the Relation Between the HI Content
of Galaxies and Metal Enrichment at their Outskirts} 

\author{Sean M.\ Moran\altaffilmark{1},
Timothy M.\ Heckman\altaffilmark{1},
Guinevere Kauffmann\altaffilmark{2},
Romeel Dav\'e\altaffilmark{3} 
Barbara Catinella\altaffilmark{2},
Jarle Brinchmann\altaffilmark{4},
Jing Wang\altaffilmark{2},
David Schiminovich\altaffilmark{5}, 
Am\'elie Saintonge\altaffilmark{6},
Javier Gracia-Carpio\altaffilmark{6},
Linda Tacconi\altaffilmark{6},
Riccardo Giovanelli\altaffilmark{7},
Martha Haynes\altaffilmark{7},
Silvia Fabello\altaffilmark{2}, 
Cameron Hummels\altaffilmark{5}, 
Jenna Lemonias\altaffilmark{5}, 
\& Ronin Wu\altaffilmark{8}
}

\altaffiltext{1}{Department of Physics and Astronomy,
  The Johns Hopkins University, 3400 N.\ Charles Street, Baltimore, MD
  21218, USA}
\email{moran@pha.jhu.edu}
\altaffiltext{2}{Max Planck Institut f\"{u}r Astrophysik,
  Karl-Schwarzschild-Str. 1, D-85741 Garching, Germany}
\altaffiltext{3}{Astronomy Department, University of Arizona, Tucson, AZ 85721, USA}
\altaffiltext{4}{Leiden Observatory, Leiden University, 2300 RA, Leiden, The Netherlands}
\altaffiltext{5}{Department of Astronomy, Columbia University, 
550 West 120th Street, New York, New York 10027, USA} 
\altaffiltext{6}{Max Planck Institut f\"{u}r Extraterrestrische Physik, 
   Giessesbach-Str., 85748 Garching, Germany}
\altaffiltext{7}{Department of Astronomy, 610 Space Sciences Building, Cornell University, Ithaca,NY 14853, USA}  
\altaffiltext{8}{Commissariat \`{a} l'Energie Atomique (CEA), 91191 Gif-sur-Yvette, France}

\begin{abstract}
We have obtained long-slit spectra of 174 star-forming galaxies
with stellar masses greater than $10^{10} M_{\odot}$  from the 
{\it GALEX} Arecibo SDSS (GASS) survey. These galaxies have both HI and
H$_2$ mass measurements. The average metallicity profile is strikingly 
flat out to $R_{90}$, the radius enclosing 90\% of the $r$-band light. 
Metallicity profiles which decline steadily with radius 
are found primarily for galaxies in our sample with low stellar mass
(Log$(M_*)<10.2$), concentration, and/or mean stellar mass density. 
Beyond $\sim R_{90}$, however, around 10 percent of
the galaxies in our sample  exhibit a sharp downturn in metallicity. 
Remarkably, we find that
the magnitude of the outer metallicity drop is well correlated with
the {\it total} HI content of the galaxy (measured as $f_{HI}=M_{HI}/M_*$).
We examine the radial profiles of stellar population ages and star formation rate 
densities, and conclude that the galaxies with largest outer
metallicity drops are actively growing their 
stellar disks, with mass doubling times across the whole disk only one third as
long as a typical GASS galaxy.
We also describe a correlation between {\em local} stellar mass 
density and metallicity, which is valid across all galaxies in our sample. 
We argue that much of the recent stellar mass growth at the edges of
these galaxies can be linked to the accretion or radial transport
of relatively pristine gas from beyond the galaxies' stellar disks.
\end{abstract}


\section{Introduction}

A proper characterization of the ages and metallicities of
the stars  in spiral galaxies, as well
as the radial dependence of the metallicity of stars and gas in these systems,
has long been recognized as a key stepping-stone to unravelling disk galaxy   
formation processes, including the roles of gas accretion, supernovae-driven 
outflows and the radial migration of stars
\citep{quirk73, tinsley80, lacey85, wyse89, kauffmann96, chiappini97, schonrich09}. 

There have, however, been rather few systematic studies of how radial 
star formation and metal abundance gradients
vary across {\em populations} of disk galaxies.
\citet{vilacostas92} carried out an analysis of
abundance gradients in a sample of 30 spiral galaxies with spectroscopy
from the literature. Barred galaxies were found to have
flatter abundance gradients than un-barred galaxies.
When the analysis was restricted to un-barred galaxies, gradients
were found to be stronger in late-type galaxies (see also
\citep[see also][]{oey93}. Vila-Costas \& Edmunds
also found the the central metallicities of spiral galaxies
were correlated with both the total stellar mass of the galaxy,
and the local central surface density of stars. 
Zaritsky, Kennicutt \& Huchra (1994) obtained uniform
HII-region based abundance gradients for a sample of 39 nearby disk galaxies.
In this study, the slopes of radial abundance gradients, when expressed in units of
dex/isophotal radius, did not exhibit any significant correlation with
luminosity or Hubble type, but it was later noted by \citet{garnett97} that gradients
expressed in dex/kpc steepen with decreasing luminosity. 
Further observations by \citet{vanzee98} focused 
on increasing the number of measurements in the outer regions of galaxies; 
they identified several galaxies where the fitted gradient changed significantly 
with the addition of higher-radius data.
Recently, \citet{moustakas10} presented resolved metallicity measurements 
for a sample of nearby
galaxies from the {\it Spitzer} Nearby Galaxies Survey \citep[SINGS,][]{kennicutt03a}, and pointed out that derived metallicity gradients  
can be sensitive to the methodology used to calibrate the strong-line
abundance estimates. 

A complete, systematic and carefully-executed survey of metallicity 
gradients in a large sample of nearby disk galaxies is urgently needed to help
put into context some recent studies that have reported unexpected or 
peculiar metallicity profiles in a number of individual galaxies. 
A number of these 
\citep[e.g.,][]{werk11, rupke10, kewley10, bresolin09}
find surprisingly flat metallicity gradients, even out to extreme 
distances, both for interacting
galaxies and peculiarly gas-rich objects, in contrast with earlier work
documenting low metallicities or large drops in the
extreme outskirts of galaxies \citep[e.g.,][]{ferguson98, kennicutt03}.
The more recent papers argue that these differences reflect 
the important role of 
gas inflows and outflows, as well as mixing, in determining how
metallicity changes as a function of radius in galaxies. 

In order to learn more about the relations between cold gas
and star formation in nearby galaxies, we  
are carrying out the {\it GALEX} Arecibo SDSS
Survey (GASS)\footnote{http://www.mpa-garching.mpg.de/GASS} \citep[][hereafter C10]{catinella10}. 
GASS is designed
to measure the neutral hydrogen content of a representative sample of $\sim 1000$
galaxies uniformly selected from the Sloan Digital Sky Survey \citep[SDSS,][]{york00} 
and {\it Galaxy Evolution Explorer} ({\it GALEX}, Martin et al. 2005) imaging
survey, with stellar masses in the range $10^{10}-10^{11.5} {\rm M}_{\odot}$ and
redshifts in the range $0.025<z<0.05$.  
GASS observations are designed to detect HI down to a gas-fraction
limit of 1.5-5\%, so the full GASS sample will be the first HI survey able to
place meaningful, unbiased constraints on the atomic gas reservoirs that 
may contribute to future growth in massive galaxies. 
We are also pursuing a companion project on the IRAM 30m
telescope, COLD GASS\footnote{http://www.mpa-garching.mpg.de/COLD$\textunderscore$GASS},
which has obtained accurate and homogeneous molecular gas masses for 
a subset of 350 galaxies from the GASS sample \citep{saintonge11}. These data 
will allow us to characterize the balance
between atomic and molecular gas in the galaxies in our sample, and
understand the  physical processes that determine how the condensed baryons
are partitioned into stars, HI and H$_2$ in the local Universe.

The third component of the GASS survey, and the subject of the current paper,
is a follow-up campaign to obtain long-slit spectroscopy for 
the COLD GASS subset of 350 galaxies. Such spectra  
allow us to link the measured gas 
contents of galaxies to their rotational dynamics, metal abundance
gradients and resolved star formation histories.
In this first paper in our series on  the GASS long-slit data, we focus 
on studying the gas-phase metallicities of star-forming 
regions as a function of radius for a partially-complete sample of 174 GASS galaxies. 

In \S2 \& 3 below, we will describe our observations, data reduction
pipeline, and our methodology for  measurement of metallicities.
In \S4, we will show how metallicity  gradients 
vary as a function of both 
global and localized galaxy properties. In \S5, we will 
present a remarkably tight relation between the total HI 
content and metallicity in the outer regions of 
galactic disks, and examine in detail a subset of galaxies exhibiting
steep metallicity drops in their outer disks.
In the following, we adopt a standard $\Lambda$CDM cosmology with
H$_0=70$km~s$^{-1}$~Mpc$^{-1}$, $\Omega_m=0.3$ and $\Omega_\Lambda=0.7$.
Stellar masses and star formation rates are calculated assuming a
Kroupa (2001) initial mass function (IMF).

\section{Observations}
GASS observations of the 21cm line of neutral hydrogen have been  obtained
with the L-band Wide receiver of the Arecibo telescope. 
Integration times are set such that we
detect any HI down to a limiting HI fraction ($f_{HI}=M_{HI}/M_*$) of 
3.5\% or less. Details of the HI sample, observations, and mass
calculations are given in C10. 

Likewise, molecular gas masses are determined through observations
of the CO ($J=1-0$) line using the IRAM 30m telescope, to similar mass
fraction limits, as part of the
COLD GASS program. Details of the observations and mass
determinations are provided in \citet{saintonge11}.

We note that both HI and CO observations provide only {\it integrated} 
measures of total atomic or molecular gas mass, respectively. Thus,
when comparing to the results from our {\it resolved} spectroscopy, it is 
important to keep in mind that our current data
gives no insight into the spatial distribution of the gas.

Long-slit spectroscopy of 174 galaxies in the GASS and COLD GASS
samples was obtained over the period 2008 October to
2010 November, using both the Blue Channel Spectrograph on the 6.5m MMT
telescope on Mt. Hopkins, AZ, and the Dual Imaging Spectrograph (DIS)
on the Apache Point 3.5m. We observe each galaxy
with the slit aligned along the major axis of the galaxy, with a slit
length that is much larger than the size of the galaxy. Our sample contains 131 galaxies
with MMT spectra, and 43 with APO spectra.

All MMT observations were obtained through a
1.25\arcsec wide slit covering the spectral range 
$\sim3900-7000$\mbox{\AA} at a spectral resolution of $\sim4$\mbox{\AA}
FWHM, equivalent to $\sigma\sim90$km~s$^{-1}$ in the rest frame of a
typical galaxy. In the spatial direction, pixels are 0.3\arcsec wide,
and we typically obtained 2x900s exposures for each galaxy.

APO observations were taken through a slightly wider, 1.5\arcsec slit,
with spatial sampling of 0.4\arcsec. Typical exposure times were somewhat longer
(2x1200s). The wavelength range of APO
spectra is more extended than MMT, covering from
$\sim3800-9000$\mbox{\AA} at a spectral resolution of 6--8~\mbox{\AA} or $\sim150$km~s$^{-1}$.

Data were reduced in IDL with routines from the
publicly available Low-Redux
package\footnote{http://www.ucolick.org/$\sim$xavier/LowRedux/} maintained
by X. Prochaska, which itself is an adaptation of the Princeton SDSS reduction
code\footnote{http://spectro.princeton.edu} to more general long-slit 
reductions. The code performs standard biasing, flat-fielding,
cosmic-ray rejection, and sky estimation on each exposure. We then
co-add the sky-subtracted exposures through a custom-written routine
that verifies and adjusts the alignment of exposures before co-addition. 
Flux calibration was achieved via observation of spectrophotometric
standards BD$+$17~4708, BD$+$33~2642, or Feige~67. 

\begin{figure*}
\includegraphics[width=2\columnwidth]{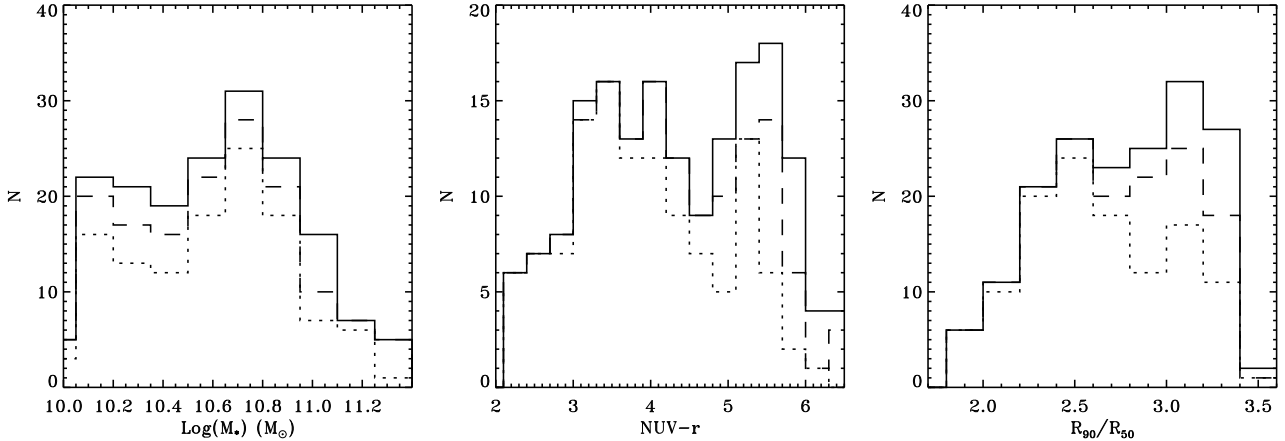}
\caption{\label{histograms} Histograms showing the distribution of
stellar mass, NUV--{\it r} colour (observed frame), and concentration index
$R_{90}/R_{50}$ for our full spectroscopic sample
(solid), complete star-forming sample (dashed), and sample with
extended ($R>0.7R_{90}$) star formation (dotted).}

\end{figure*}

APO spectra are recorded as separate files for the  blue and the red channels; 
these are
reduced independently, as described above, and then joined.  To eliminate
any inconsistency in the flux scaling between the two channels,
we rescale the spectra as follows: 
we select a 75\mbox{\AA} window near 5450\mbox{\AA} where 
the dichroic element that splits 
the incoming flux has non-zero transmission into
both channels. Within this window, we measure the 
flux normalization of each spectrum, calculate the average of the two, 
and place both  on a uniform flux scale by applying 
the average scaling to both  spectra. The blue and the
red spectra can then be joined through a 
simple concatenation, averaging values within the small window of overlap.

Thanks to existing, very accurately calibrated SDSS photometry 
and spectra, we are able to
refine our flux calibrations (for both MMT and concatenated APO spectra) by matching 
to SDSS. First, we convolve our spatially integrated spectra with the response functions
of the SDSS {\it g} and {\it r} filters. 
We next measure {\it g} and {\it r} magnitudes directly on the SDSS images
through an aperture matched to our slit, following the procedure of \citet{wang11a}, 
described further in \S3. 
We then apply the average difference in $g$ and $r$-band magnitudes
as a scalar correction factor to our spectrum. 
Such corrections allow us to accurately flux calibrate spectra that
were taken in non-photometric conditions. Typical correction factors are $\sim0.25$~mag
for observations taken through light cloud cover. 
Typically the corrections derived from $g$ band alone versus those
for $r$ differ by less than
0.02~mag, verifying that our {\it relative} spectrophotometry is quite
accurate.
When we
compare SDSS fiber spectra to the central portion of our slit spectra, 
matched as well as possible to the SDSS 3\arcsec aperture, 
the resulting spectrophotometry 
agrees with SDSS to better than 10\% across the full wavelength range.

\section{Analysis}

In this paper, we focus on a subset of 151 galaxies (out of the
174 galaxies in the GASS spectroscopic sample)
where we have detected significant emission lines ($>3\sigma$) 
from star forming regions     
anywhere within the area probed by our
spectroscopic slit. We exclude regions contaminated by
AGN emission (see below). 
The presence of emission lines is necessary for measuring 
gas-phase metallicities. Within the star-forming subset of
151 galaxies, 119 have star formation that can be
traced to at least $2R_{50}\sim 0.7R_{90}$, where
$R_{50}$ and $R_{90}$   are the radii enclosing 50\% and 
90\% of the galaxy's r-band light, respectively. 
In \S5, which discusses metallicities in the 
outer regions of galaxies, the galaxies are drawn from this subsample.

In Figure~\ref{histograms}, we show how these various subsamples are distributed in 
stellar mass, NUV--{\it r} color (observed frame), and concentration index
($R_{90}/R_{50}$). We note first that the 174-galaxy spectroscopic sample 
has the same distribution of galaxy properties as the full GASS sample of 1000
galaxies.
Both are selected to have a flat  stellar mass distribution ( left histogram
in Figure~\ref{histograms}).  Further
restricting the sample to, first, star-forming galaxies, and second, galaxies
with star formation at  {\it large  radius}, serves only to exclude a number
of red, high concentration galaxies. 
We note that galaxies across the stellar mass range
$10<Log(M_*)<11$ are still well sampled despite these cuts.

We first follow the procedure outlined in \citet{moran10}
to determine galaxy rotation curves, which allows us to correct the
spectrum from each individual row to a common rest frame before binning. This helps
to avoid velocity blurring of the coadded spectrum. (Note that we will not
otherwise discuss galaxy rotation curves and dynamics in this paper).
For each two-dimensional spectrum sampled at 0.3 or 0.4 arcsecond
resolution, we then
perform adaptive binning in the spatial direction to ensure an adequate S/N in
each spatial bin in our analysis. 
Beginning at the galaxy center, defined as the spatial row with peak
flux, we coadd bins working outward one row at a time, stopping when we reach
an integrated S/N of 15 (per Angstrom). If S/N does not reach 15 by the time
3\arcsec of slit has been coadded, then we stop at that width so long
as S/N is greater than 10, and begin a new bin. 
We proceed outwards in this manner,  with three
additional breakpoints where the $S/N$ requirements
are changed: we require $S/N>10$ up to size 4.5\arcsec, followed by $S/N>8$
up to size 6\arcsec, and $S/N>6$ at distances beyond that. 
The procedure terminates when  further binning fails to add to
either the continuum S/N or the S/N in the H$\alpha$ line, which is 
tracked to ensure that we do not discard any faint emission.
We note, also, that all bins are constrained to be at least as long as
the slit width (1.25\arcsec or 1.5\arcsec), since finer binning 
does not provide added spatial information.

The end result of this process is a set of spectra for each galaxy
where bin size generally increases with radius, but the S/N in each
bin falls much more slowly with radius 
than it would if we had used uniform bin sizes.

We employ a modified version of the technique of 
\citet[][hereafter T04]{tremonti04}
to measure the strengths of key emission and absorption lines in each
spatial bin for each galaxy, as described in  detail in \citet{brinchmann04}
and \citet{moran10}. 
In brief, we fit each spectrum to a
linear combination of templates drawn from \citet{bc03} single
stellar population models of varying metallicities. 
Next, we subtract the
best-fitting stellar continuum model from the measured spectrum,
creating an emission-line only spectrum where the Balmer emission lines can
be measured free of contamination from the underlying stellar absorption.

For most spatial bins, our code fits for the best-match velocity dispersion
of the continuum spectrum, but for bins with marginal S/N in the continuum
($3<S/N<8$~per Angstrom), we instead adopt the median velocity dispersion of
all the high S/N bins from the same galaxy.

\begin{figure}[t]
\includegraphics[width=\columnwidth]{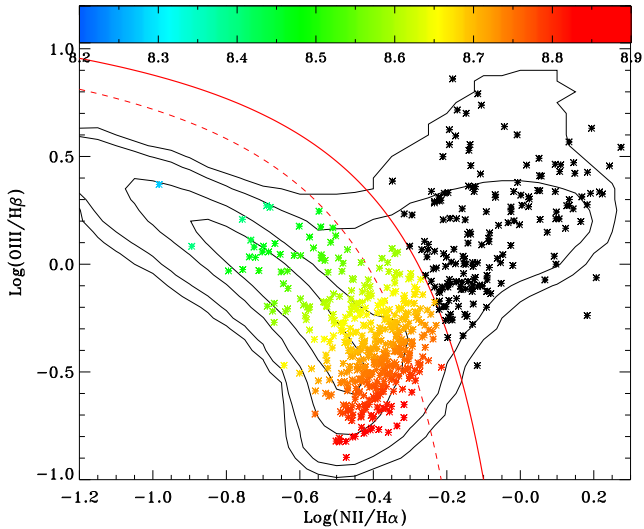}
\caption{\label{bpt}
Line ratio diagram for 
all independent spatial bins with significant ($>5\sigma$) emission in
all four lines, drawn
from all 174 spectra in our sample.  
Colored points indicate galaxy regions of varying metallicities, as indicated in the
color bar at top  of the figure.  
Emission-line regions with line ratios consistent with excitation
dominated by an AGN are plotted as black points, and lie above the 
solid red line. Between the solid and dotted red lines, points may 
reflect a small AGN contribution to the flux. 
Contours show where 33\%, 67\%, 95\% and 99\% of SDSS galaxies
with redshifts in the GASS range ($0.025<z<0.05$) lie. Note that for the SDSS galaxies,
line ratios are measured within the fiber.}
\end{figure}

In cases where only nebular emission is detected, common at the
outskirts of some galaxies, no continuum fitting
is possible,  so we instead fit a low-order polynomial to the
spectrum to correct for small imperfections in our sky
subtraction, which arise  when co-adding across a large
portion of the slit. We then measure emission lines using the  
polynomial-subtracted spectrum. 

We fit a Gaussian function to the emission lines, with the width of the Gaussian constrained
to a single value for all lines in a given spectrum. As the binned
spectra have already had any galactic rotation component removed
before coaddition, the choice of a  
single width for all lines is a reasonable first approximation. The  
positions of the line centroids  are constrained to their
rest-wavelengths.  In addition to  the Balmer lines H$\alpha$, H$\beta$, 
and H$\gamma$, used to
estimate dust extinction and star formation rate (see below), we also
measure the forbidden lines [\ion{O}{2}]~$\lambda$5007, [\ion{N}{2}]~$\lambda\lambda$6548,6584, and,
when possible, [\ion{O}{2}]~$\lambda\lambda$3726,3729 and [\ion{S}{2}]~$\lambda\lambda$6717,6731,
which are required to  measure metallicity across each galaxy.

We estimate the dust extinction within the nebular gas by calculating the
Balmer decrement, which we  define as  the ratio of H$\alpha$/H$\beta$ to the
case B recombination ratio of 2.87 \citep{osterbrock89}. We use the 
formula $E(B-V)_{gas}=1.97 \log({\rm H}\alpha/{\rm H}\beta/2.87)$, where we
adopt the \citet{calzetti00} attenuation curve with $R_V'=4.05$. 
We note that all SFRs reported below have been calculated after
correcting fluxes for extinction using the Balmer
decrement. 

After correcting H$\alpha$ luminosities for dust, we measure star
formation rates using  the equation in  \citet{meurer09}: 
SFR ($M_\odot$~yr$^{-1}$)=L$_{{\rm H}\alpha}/$($6.93\times
10^{33}$~W), corrected to a \citet{kroupa01} IMF in order to be
consistent with SFRs reported in the SDSS catalogs. 
Then, by dividing by the area of the galaxy
covered by each  portion of the slit (e.g., for MMT, $1.25\arcsec \times
0.3\arcsec N$, where N is the number of individual rows that went into
each co-added spectrum), we estimate the SFR surface density as a function of
position across each galaxy.

Stellar mass densities ($\mu_*$) are derived for each individual spectral bin
by re-measuring SDSS photometry through an aperture matched to each slit segment,
and calculating stellar mass for that segment using  the SED
fitting procedure described in \citet{wang11a}. Dividing by the area under the slit
then yields  $\mu_*$, and we further calculate a local measure of specific
star formation rate, sSFR, under each slit segment by dividing SFR surface density
by these $\mu_*$ values. 

We limit our analysis of stellar absorption features in this work to the 
$D4000_n$ index, which measures the
strength of the 4000\mbox{\AA} break \citep{balogh99}. This index is  an indicator of stellar
population age \citep[see, e.g.][]{kauffmann03a}. We measure this
index directly from  the spectrum, but only include measurements from
spatial bins where the continuum S/N around 4000~\mbox{\AA} is $>5$
(per \mbox{\AA}).

When we compare our results with those derived from SDSS
fiber spectra, we select all galaxies in SDSS Data Release 7 \citep[][hereafter, DR7]{abazajian09} with
redshifts in the GASS range ($0.025<z<0.05$). We make use of star 
formation rates, stellar masses, and metallicities provided in  the MPA-JHU value added 
catalogs\footnote{http://www.mpa-garching.mpg.de/SDSS/DR7/}
We also use raw line fluxes from these catalogs for calculating metallicities 
as described in the next section. 

We note that measurements of emission and absorption line strengths for all
galaxies in the GASS spectroscopic sample will be made available 
online\footnote{http://www.mpa-garching.mpg.de/GASS/} upon
completion of the survey.

\subsection{Metallicities}
To estimate gas-phase metallicities, we rely primarily on the O3N2 empirical
index described by \citet{pp04}. This index relates two
line ratios, [\ion{N}{2}]~$\lambda$6583/H$\alpha$ and [\ion{O}{3}]~$\lambda$5007/H$\beta$ to the
metallicity of the ionized gas.
All four of these emission lines fall within
our spectral coverage for both MMT and APO spectra. As for the star
formation rates, we first correct all four lines for
extinction. However, because the index relies on ratios of lines that
are quite near each other in wavelength, their sensitivity to
extinction is negligible. We express gas-phase metallicities 
in the usual manner here, in terms of $12+Log(O/H)$. 
The \citet{pp04} relation is $12+Log(O/H)=8.73-0.32\times
O3N2$, where $O3N2=Log(([$\ion{O}{3}$]/H\beta) / ([$\ion{N}{2}$]/H\alpha)$. 
This relation is valid at least across the range
$8.0<12+Log(O/H)<9.0$, which encompasses all of our 
measured metallicities.

The four lines that go into the O3N2 index are the same used in the
classic Baldwin, Phillips \& Terlovich (1981), or BPT, 
diagram used to discriminate emission from AGN from that due
to star-formation. In Figure~\ref{bpt}, we plot line ratios for  all
spatial bins from our sample, where all four lines are measured with $S/N>5$.
Points are color-coded according to the metallicity estimated using
the O3N2 indicator, as illustrated by the color bar. The dashed line
shows the division between galaxies with star-formation dominated and 
AGN-dominated or AGN/star-formation composite emission defined  by \citet{kauffmann03b}.

\begin{figure*}
\includegraphics[width=\columnwidth]{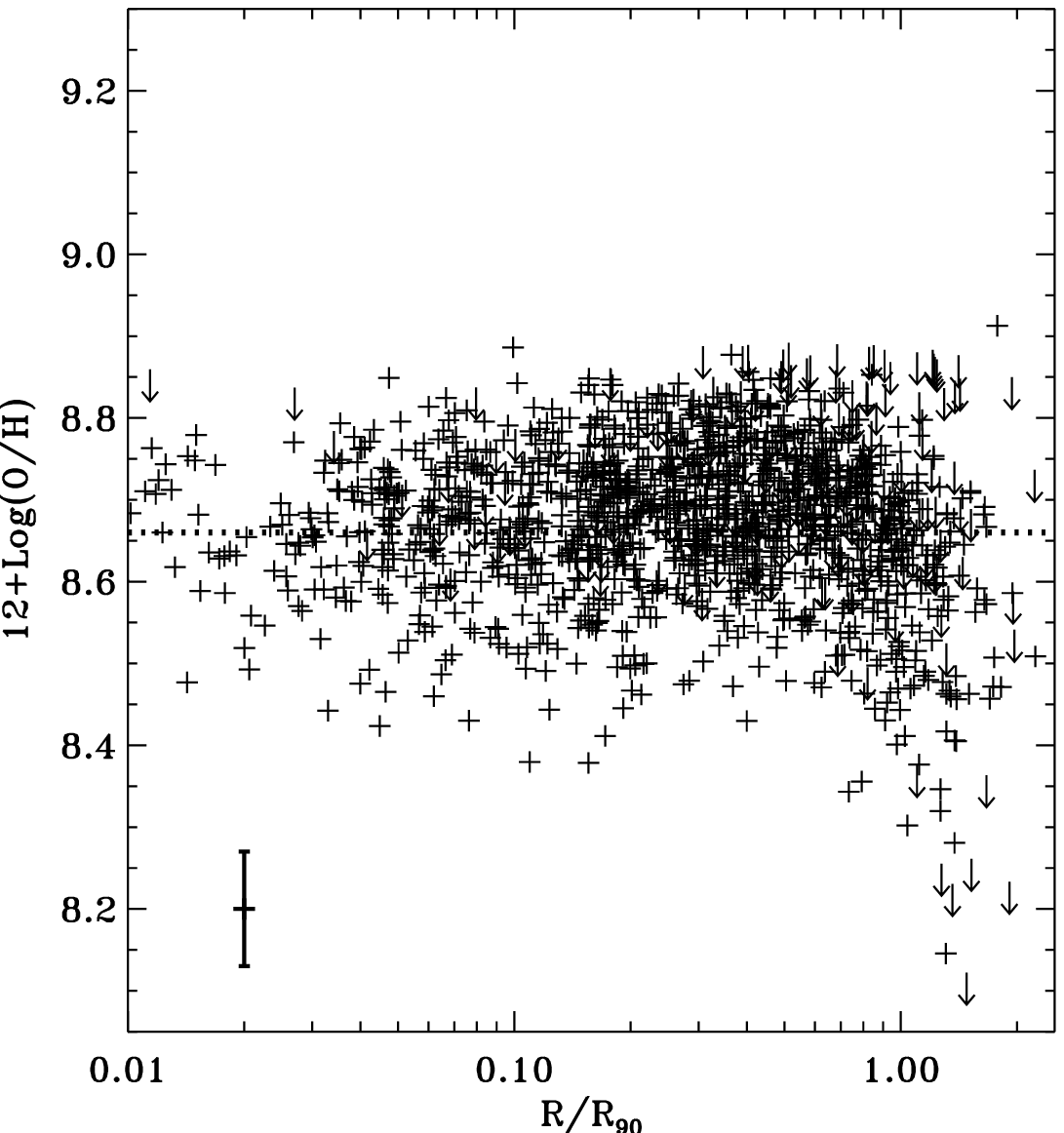}
\includegraphics[width=\columnwidth]{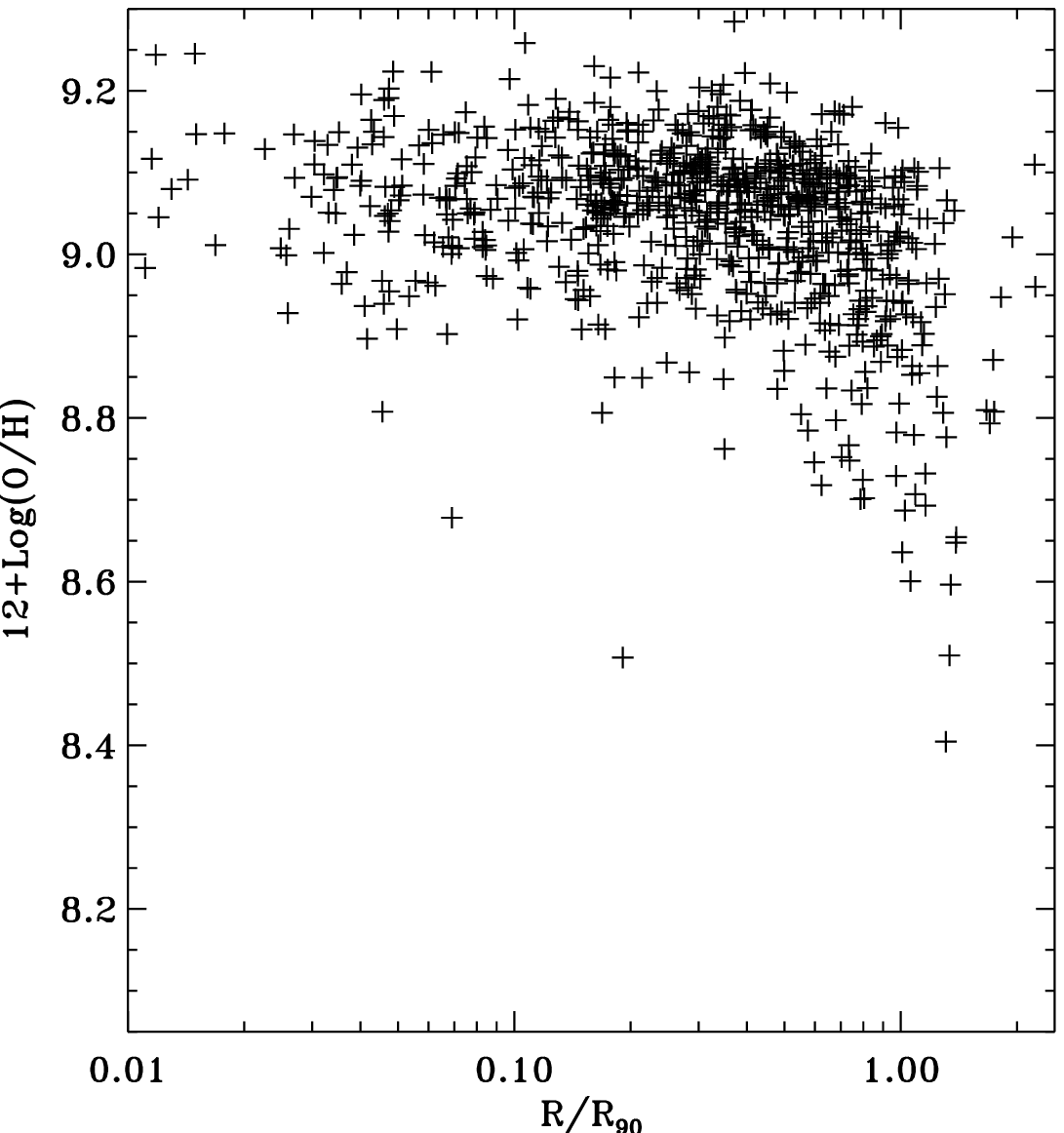}
\caption{\label{metal_r90} Gas-phase metallicities as a function of radius, 
$R/R_{90}$, for all spatial
bins in all galaxies in our sample. Metallicities are
estimated from the Pettini \& Pagel (2004) O3N2 index (left), and 
from the method of T04 (right).
Black crosses show metallicities with nominal 1-$\sigma$ statistical uncertainties of $<0.15$~dex (left, $n=1298$) 
or $<0.18$~dex (right, $n=828$), respectively. In the left-hand panel, downward-pointing arrows denote upper limits
in cases where not all emission lines were measurable
($n=182$). Our typical uncertainty of 0.07~dex is illustrated by the 
error bar at lower left.} 
\end{figure*}

We exclude from our analysis any regions where the measured emission
exhibits AGN-like line ratios, which are marked in black in Figure~\ref{bpt}.
Note that we choose a dividing line that is slightly offset ($+0.1$~dex in both axes) 
from that used by \citet{kauffmann03b}, 
shown as a thick solid line in Figure~\ref{bpt}. This is done
in order to avoid throwing out star-forming points that fall just over the line 
due to statistical fluctuations. We thus include a small number of points
with so-called `composite' AGN and star-formation emission lines. These make
up less than 10\% of the total, and the fractional 
AGN contamination for points
close to the star-forming sequence of the BPT diagram is low \citep{brinchmann04}, 
so their inclusion is not likely
to bias our metallicity or star formation estimates in any significant way.

We note that the number of AGN points marked in Figure~\ref{bpt} is
larger than the number of galaxies in our sample (180 points c.f. 174 galaxies).
This is because AGN emission often extends 
into more than one spatial bin at the centers of
our galaxies, where our bins tend to be only slightly larger than the typical 
seeing disk. In addition, a number of the non-star-forming galaxies in our sample
exhibit faint, spatially extended emission with LINER-like line ratios, and these
are also included on the diagram.
In total, 47\% of our galaxies exhibit central emission
indicative of an AGN, in line with expectations for a population in
this mass range \citep{kauffmann03b}. Approximately 10\% of our galaxies
have extended emission (beyond 3\arcsec from the nucleus) 
with LINER-like line ratios.

At the lowest metallicities, it is clear from Figure~\ref{bpt} that
[\ion{O}{3}]/H$\beta$ varies only slightly, and so the [\ion{N}{2}]/H$\alpha$ ratio alone is 
sufficient to determine metallicity.
Likewise, in the intermediate metallicity regime, 
a lower limit on the [\ion{O}{3}]/H$\beta$ ratio sets an upper limit
on metallicity even when H$\beta$ is undetected.
Spatial bins  with  upper limits on metallicity make up approximately 14\% of the sample,
and will be clearly marked in any figures where they are included.

To ensure the reliability of our O3N2-based metallicities, we have also
calculated metallicities using the method described in T04. 
This Bayesian technique determines the most 
likely metallicity using information from all available  emission
lines, and so utilizes  lines beyond the four considered
so far, including the [\ion{O}{2}]~$\lambda$3727 doublet and the [\ion{S}{2}]~$\lambda\lambda$6717,6731
lines. As will be discussed in
more detail in \S4, we find that metallicities calculated in these
two ways yield very similar results for metallicity gradients and
their trends as a function of galaxy mass and type.

Despite the well-known disagreements in the overall metallicity scale
exhibited by these and other metallicity indicators, recent
work by \citet{kewley08} provides a series of tabulated
functions that allows us to convert metallicities measured with one
method onto the metallicity scale of another. To
estimate the typical uncertainties on our metallicity measurements,
we therefore convert our T04-style metallicities onto the O3N2 scale
in order to make a direct comparison. We find that the rms difference
in metallicities measured according to these two methods is
$\sim$0.07~dex, which is comparable to the agreement quoted by \citet{kewley08}. 
Except for metallicity upper limits and a
small number with low S/N ($\sim10\%$), 
the formal statistical uncertainties
on our measurements are typically much smaller than 0.07~dex, with a
median of only 0.01~dex. We will thus adopt 0.07 as a reasonable
estimate of the uncertainty in all of our metallicity measurements.

For the subset of galaxies
observed with the Apache Point telescope, we have access to a broader
spectral range which allowed us to compare metallicities derived from
O3N2 to those derived using the R23 index \citep[following][]{kk04}  
which utilizes [\ion{O}{2}]~$\lambda$3727 in addition
to [\ion{O}{3}]~$\lambda$5007. After adjusting for systematic differences between the
two metallicity scales \citep{kewley08}, we find that the two indicators
are in agreement with an rms difference  of approximately
0.15 dex.

Finally, we note that variations in the N/O ratio in galaxies have been
shown to bias both O3N2 \citep{perezmontero09} and T04 \citep{yin07} 
metallicities, in the sense that emission line regions with high N/O
ratios tend to have their metallicities overestimated compared to
so-called direct-$T_e$ based metallicities. However, this bias sets in 
at different thresholds in N/O for O3N2 compared to T04. Thus, if 
elevated N/O ratios were a significant source of bias in our sample,
we would expect to see divergence in the O3N2 and T04 metallicities for
some fraction of points where the N/O ratio is in between the two
thresholds. Given the very good agreement of 0.07~dex mentioned above, 
we conclude that variations in N/O ratio are not a significant source 
of bias in our sample.

\section{Results}

\begin{figure*}[t]
\centering
\includegraphics[width=2\columnwidth]{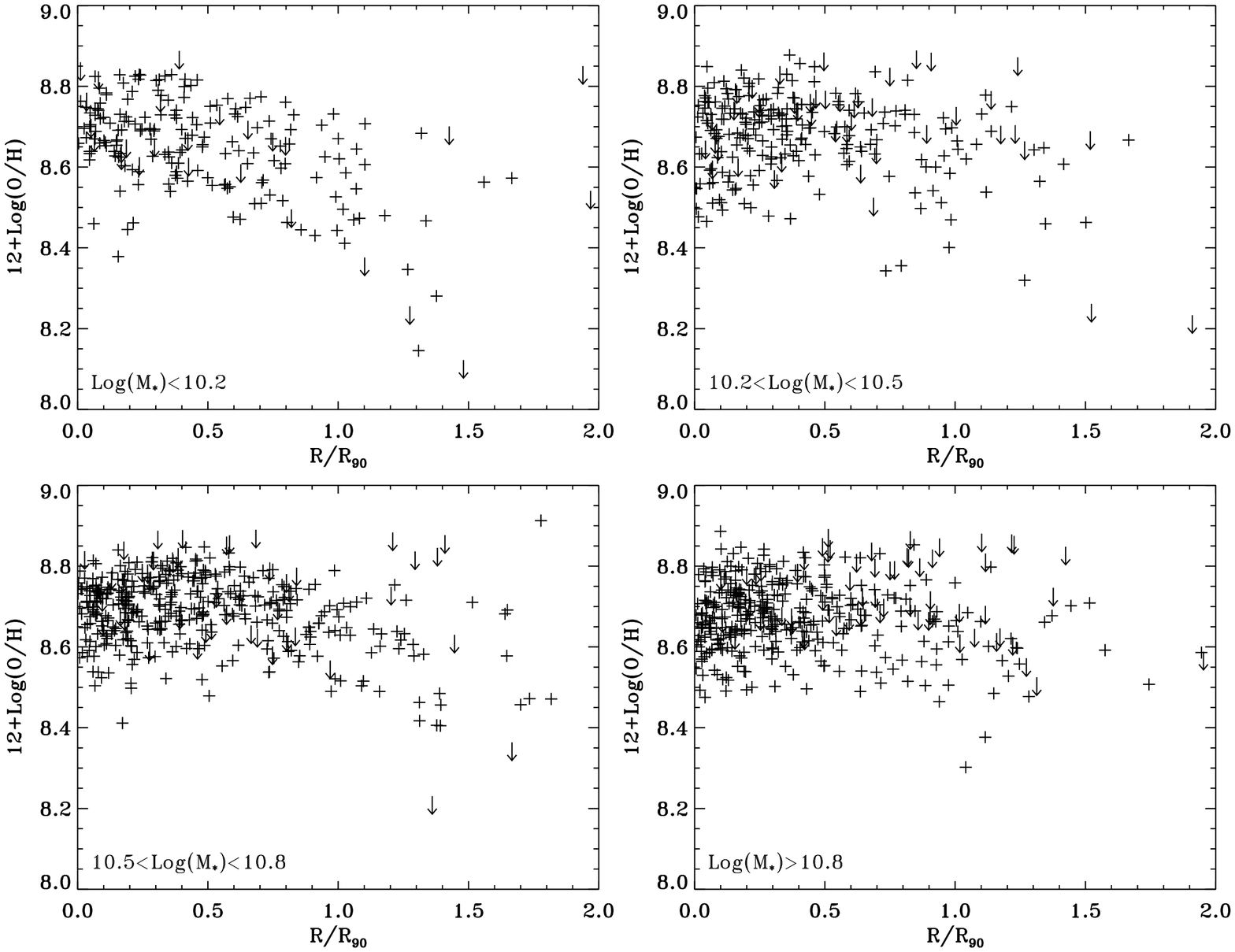}
\caption{\label{metal_by_mass} Metallicity versus radius ($R/R_{90}$)
for all spatial points divided into four bins of galaxy stellar mass, 
as marked below each plot. While low metallicity points in the outer disks
are found in all bins, only galaxies with Log$(M_*)<10.2$ exhibit metallicities 
that clearly decline  with radius on average.
}
\end{figure*}

In the left panel of Figure~\ref{metal_r90}, 
O3N2-derived metallicities for all 1298 
measured points (excluding AGN) in the 151 galaxies in our
sample of star-forming galaxies are plotted as a function of radius 
normalized by $R_{90}$ ($R/R_{90}$). We adopt a logarithmic x-axis for
clarity, since our adaptive binning technique yields a large number of
narrow bins at small radius and a comparatively smaller number of wide
bins at high radius. We include all measured metallicities 
with nominal 1-$\sigma$ statistical uncertainties of $<0.15$~dex. As 
discussed above, only $\sim10\%$ of these points have statistical uncertainty 
higher than our typical systematic uncertainty of 0.07~dex. 
Downward pointing arrows indicate upper limits
for 182 points where not all emission lines were measurable.
In the right-hand panel, we plot metallicities determined via the T04 method, this time
including all 828 points with an uncertainty from the Bayesian analysis of
$<0.18$~dex.

The large systematic offset between metallicity indicators is dramatically clear from 
Figure~\ref{metal_r90}, but the difference in scales is as expected from the
conversion formulae given in \citet{kewley08}. Of more importance are the
two key features that both panels of Figure~\ref{metal_r90} have in {\it common}.
First, we note that the vast majority of points are clustered in a narrow range of metallicity, 
which on the O3N2 metallicity scale is very near the solar value of $12+Log(O/H)=8.66$ \citep[][marked 
as a dotted line]{asplund04}. That is, the radial profiles appear quite flat across most of their range.
Second, both panels of Figure~\ref{metal_r90} feature a significant drop in metallicity 
for some points located at large radius ($R>0.7R_{90}$). This means that in some galaxies, 
gas-phase metallicity drops precipitously as one approaches the edge of the visible galactic disk.
Thus, although the absolute metallicity scale differs quite a lot---note, in particular,
that most galaxies appear to feature metallicities $>2\times$ solar abundance on the T04 scale
---we find that the {\it relative} metallicity profiles derived using both 
methods are very similar. In the following, we will examine both key features of these profiles
in more detail.

\subsection{Inner Metallicity Profiles}
Let us first consider the inner metallicity profiles ($R<R_{90}$); we will 
discuss the outer metallicity
drops in more detail below. The fact that metallicities appear essentially flat out
to nearly $R_{90}$ for the GASS sample may seem somewhat surprising. Previous work
on galaxy metallicity gradients \citep{zaritsky94, moustakas10} feature
a diverse array of gradients, with some being flat, but many others exhibiting steady
declines toward higher radius. Though their total sample sizes are
smaller, and their galaxies are selected differently from ours, 
we might expect to see at least some galaxies with declining gradients.

To examine whether plotting all points on a single plot might be 
obscuring the presence of a sub-population of galaxies
with declining metallicities,  Figure~\ref{metal_by_mass}
shows metallicity as a function of radius divided into four bins of  
host galaxy stellar mass. We now use a linear
rather than a logarithmic scale in $R/R_{90}$ in order to
highlight trends in the inner region of the galaxy.
It seems that galaxies in the lowest mass bin {\it do}, on average, 
exhibit significant metallicity gradients, where the Spearman's rank 
correlation coefficient, $\rho$, between $R/R_{90}$ and $12+Log(O/H)$ 
is modest at $|\rho|=0.34$, but still significant at the $5\sigma$ level. 
In contrast, metallicities in the higher mass bins appear to 
be quite flat as a function of radius, and show no 
statistically significant deviation from zero slope. 

In the Appendix,
we plot the individual radial profiles of the 100 galaxies with $\ge 8$ 
measured points, arranged in order of ascending stellar mass. From these,
one can verify that most inner gradients are flat, but a minority do
have sloping gradients. The distribution in stellar mass of these sloping
gradients is hard to discern from the plots, but we can attempt to quantify 
this further.

\begin{figure*}[t]
\includegraphics[width=0.66\columnwidth]{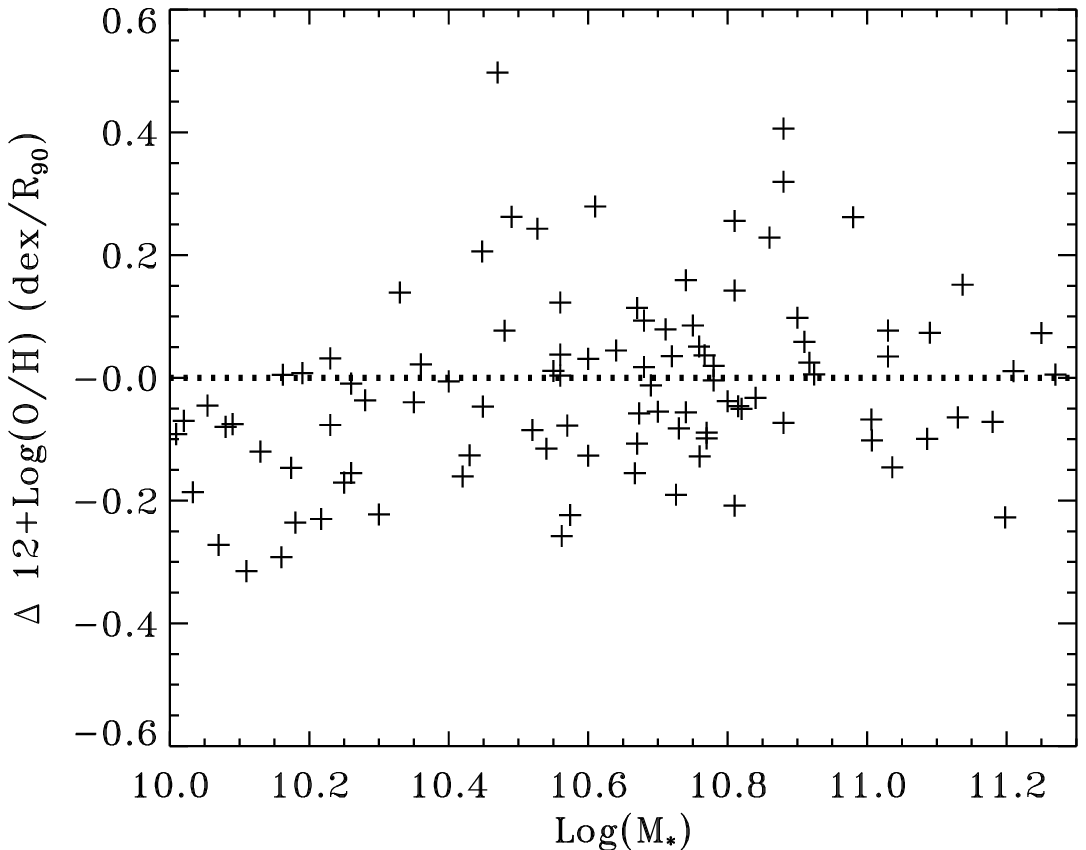}
\includegraphics[width=0.66\columnwidth]{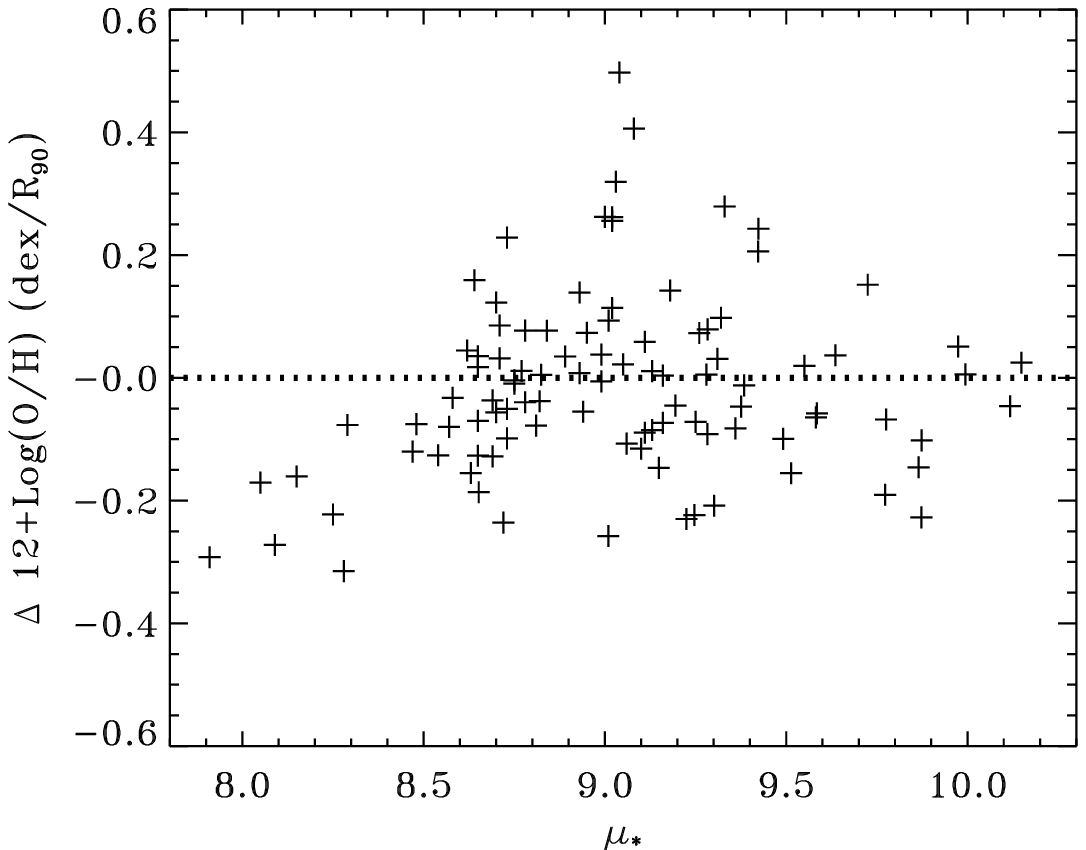}
\includegraphics[width=0.66\columnwidth]{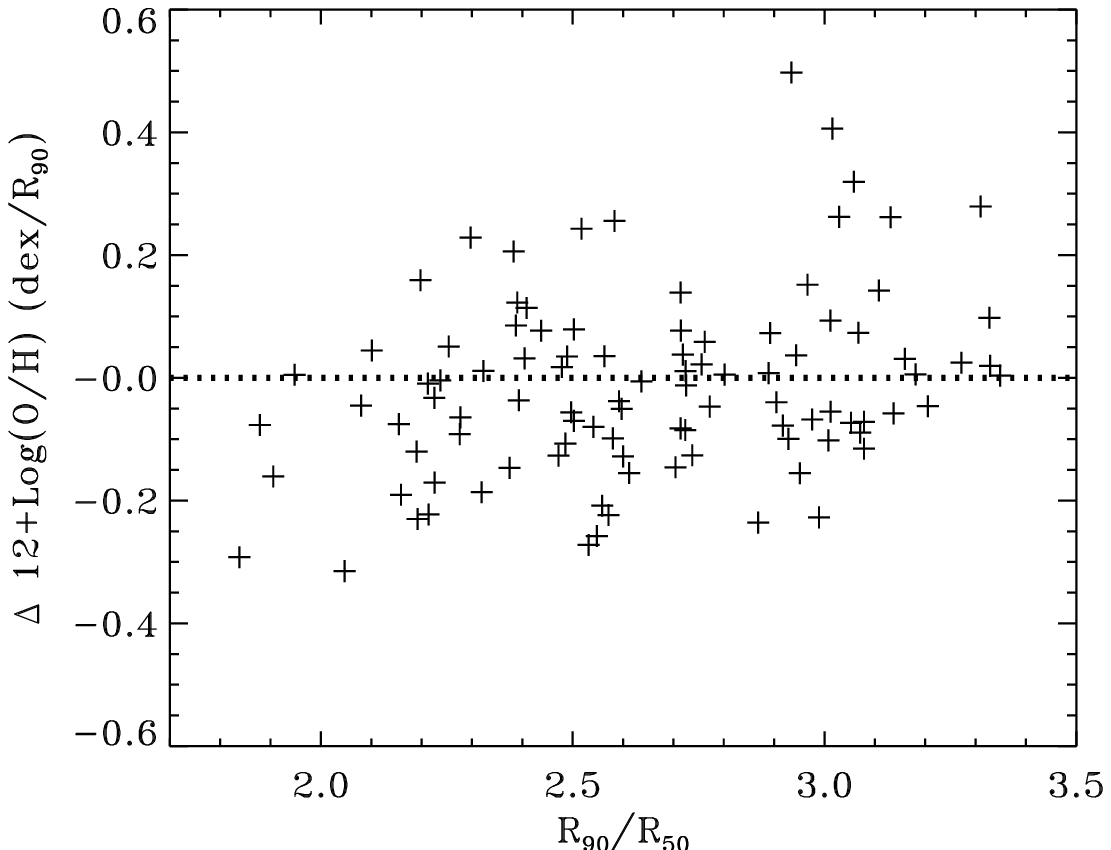}
\caption{\label{metal_gradient} Metallicity gradient in terms of dex per
$R_{90}$ for each of 100 galaxies with greater than 8 measured points, plotted
versus stellar mass (left), average stellar mass density (middle), and
concentration (right). Gradients are measured using a linear least-squares bisector fit.
Uncertainty is estimated to be 0.15 dex, and so
galaxies with significantly declining gradients occur predominantly at the 
lowest masses, mass densities, and concentrations in the GASS sample.
}
\end{figure*}

\begin{figure}[t]
\includegraphics[width=\columnwidth]{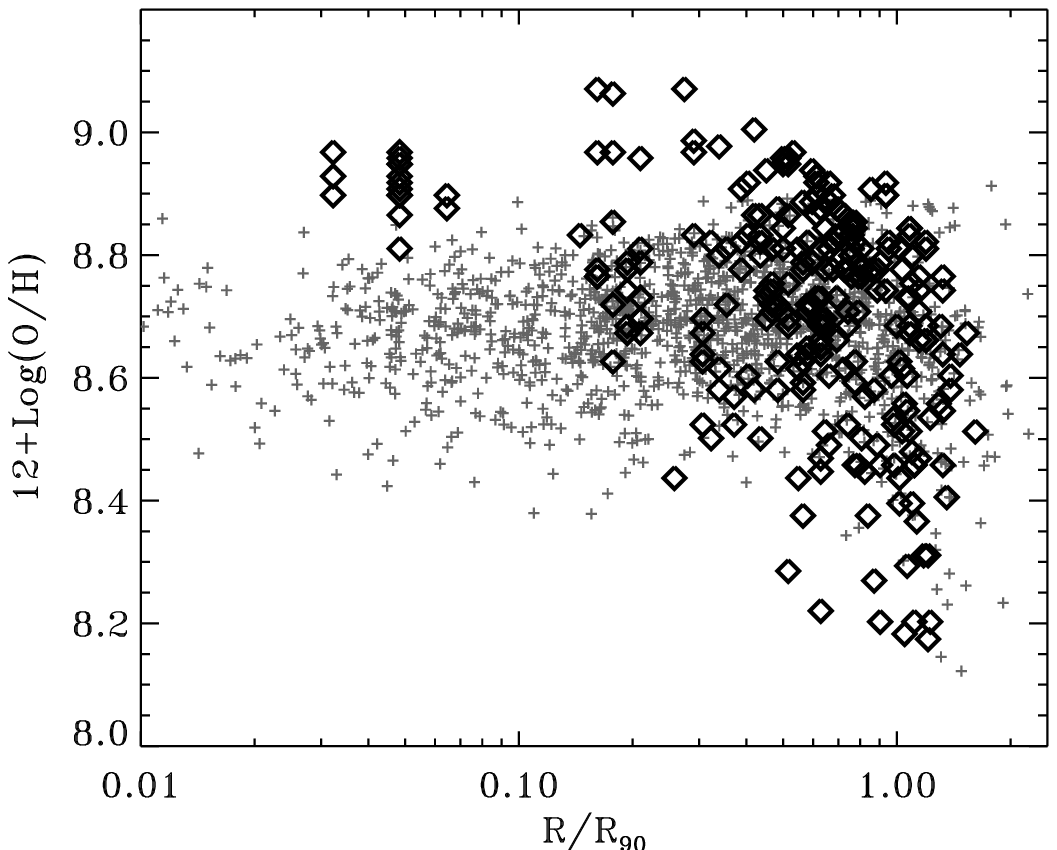}
\caption{\label{sings_compare} Metallicity versus $R/R_{90}$ for the
  massive (Log($M_*>10$)) galaxies in the SINGS sample (open
  diamonds), compiled by \citet{moustakas10}, including only
  galaxies with $>8$ measured metallicity points. R23 metallicities
  have been converted to the O3N2 scale using the formula in \citet{kewley08}, 
  and radii reported in terms of the $D_{25}$ radius have
  been scaled to an $R/R_{90}$ scale assuming the median relation
  $R_{90}=0.6*(D_{25}/2.)$. GASS points are plotted underneath as
  grey crosses. 
}
\end{figure}

In general we prefer to avoid quoting gradients in terms 
of a single number, e.g., dex of decline
per $R_{90}$, because  not all profiles are well fit by a straight line. 
Nevertheless, we have performed
these crude linear fits, and plot in Figure~\ref{metal_gradient} the best-fit slope 
to each galaxy gradient as a function of stellar mass, stellar mass density ($\mu_*$), and
concentration ($R_{90}/R_{50}$), limited to the
100 galaxies shown in the Appendix. We plot slopes
derived from a linear least-squares bisector fit. We used the IDL astronomy library 
routine SIXLIN, which performs fits with  six different
linear regression methods \citep{isobe90},
to examine  how robust the slopes are to 
varying the fitting method. We find that slopes derived using  
different methods have typical rms agreements of only
$\pm0.15$~dex$/R_{90}$, which is comparable to the scatter in the derived gradients
at fixed $M_*$ shown in Figure~\ref{metal_gradient}. 

Despite the high uncertainty, it seems that
most of the objects with plausibly declining
gradients ($\le-0.1$ to $-0.2$ dex per $R_{90}$) occur in the lowest 
stellar mass galaxies in the GASS sample, consistent with the average
profile in Figure~\ref{metal_by_mass}. Since low-mass galaxies
also tend to have low stellar mass densities and concentrations, it
is not surprising that declining gradients also occur preferentially 
at low $\mu_*$ and concentration. The formal Spearman rank correlations are
small ($|\rho|\sim0.3$) but statistically significant for stellar mass and concentration,
while that for $\mu_*$ is marginal. 

Bar fractions have also been measured for face-on galaxies
within the GASS sample \citep{wang11b}, but only a small number of 
galaxies studied here ($\sim10$) unambiguously contain bars; 
those that we identify do not show any clear difference in metallicity 
gradient from the overall sample, but better statistics will be needed
to draw any meaningful conclusions.

While we are unable to say whether morphological 
characteristics (e.g., concentration) or galaxy mass is more important 
in setting the slope of galaxies' metallicity gradient, our results
nevertheless suggest that declining
gradients are more frequently found in galaxies below some 
threshold in stellar mass 
and/or concentration (i.e., late-type morphology), which is consistent
with the results of \citet{vilacostas92}. 

To return to our original question, then, could this apparent threshold 
explain why earlier samples seemingly contain a higher incidence of 
galaxies with declining gradients?
To check, we have re-examined the 21 SINGS sample galaxies
with metallicity gradients presented in \citet{moustakas10}. By eye
examination reveals  that
8 have unarguably declining gradients,  7 have seemingly flat gradients, with the remainder
ambiguous. We then estimate the stellar masses of these galaxies 
via a simple scaling of their K-band
luminosities by a mass-to-light ratio set by their {\it B--V} colors ($Log(M/L_K)=-0.56+0.135(B-V)$) \citep{bell03}, corrected to our Kroupa IMF. 
We find that 5 out of 8 galaxies with declining gradients 
have Log$(M_*)<10.2$, while
5 out of 7 with flat gradients lie above this mass. Similarly, the most luminous galaxies
in the sample of \citet{garnett97} appear to have quite flat gradients as well.

This is suggestive, but we would ideally like to compare gradients from the
literature more directly with our sample, in particular for galaxies falling
within the GASS mass range.
\citet{moustakas10} have made available electronically a
compilation of points with metallicity measured on the R23
system. We have obtained this data, and we use the \citet{kewley08} 
formula to convert to the O3N2 system. \citet{moustakas10} recorded the positions of
individual metallicity points in terms of the $D_{25}$ optical radius,
which differs from our chosen $R_{90}$. 
We use the subsample with SDSS coverage to estimate that
$R_{90}=0.6*(D_{25}/2.)$.

\begin{figure*}[t]
\includegraphics[width=\columnwidth]{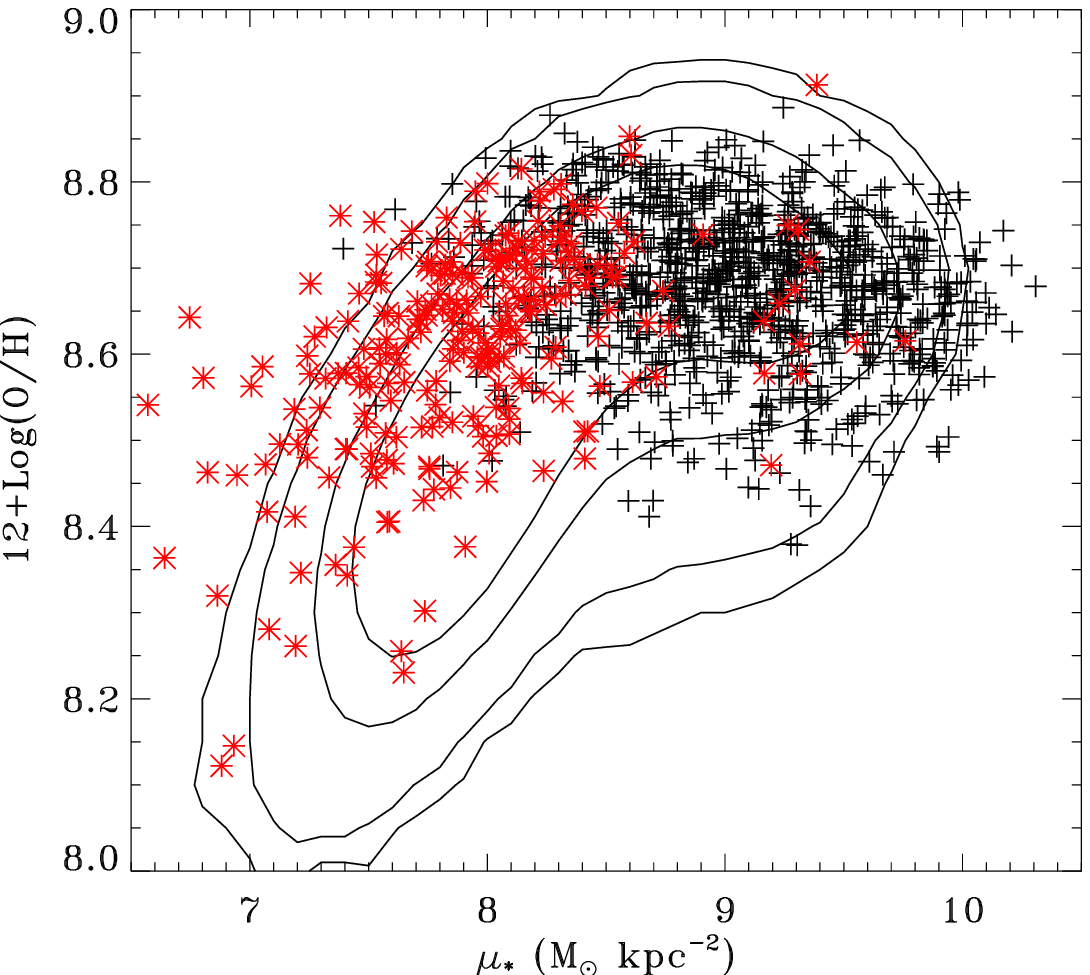}
\includegraphics[width=\columnwidth]{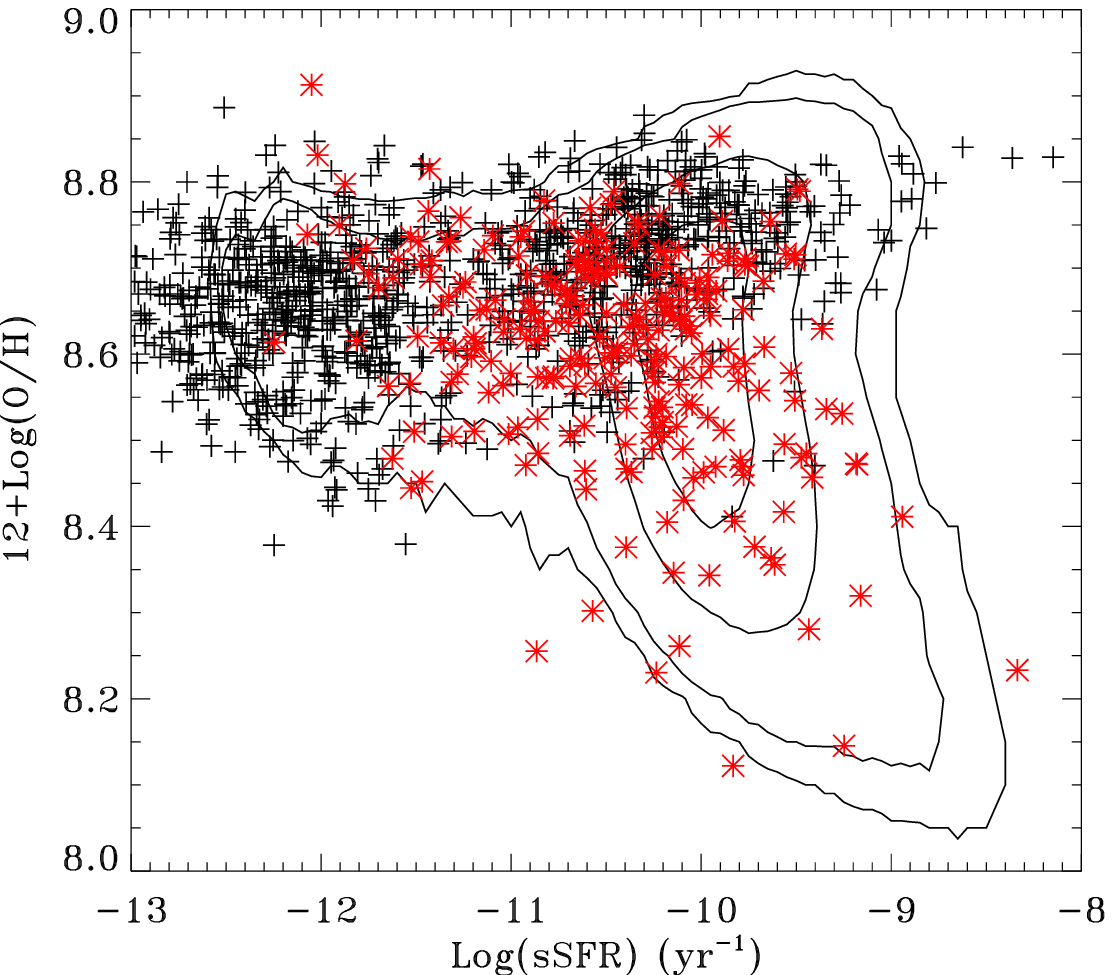}
\\
\includegraphics[width=\columnwidth]{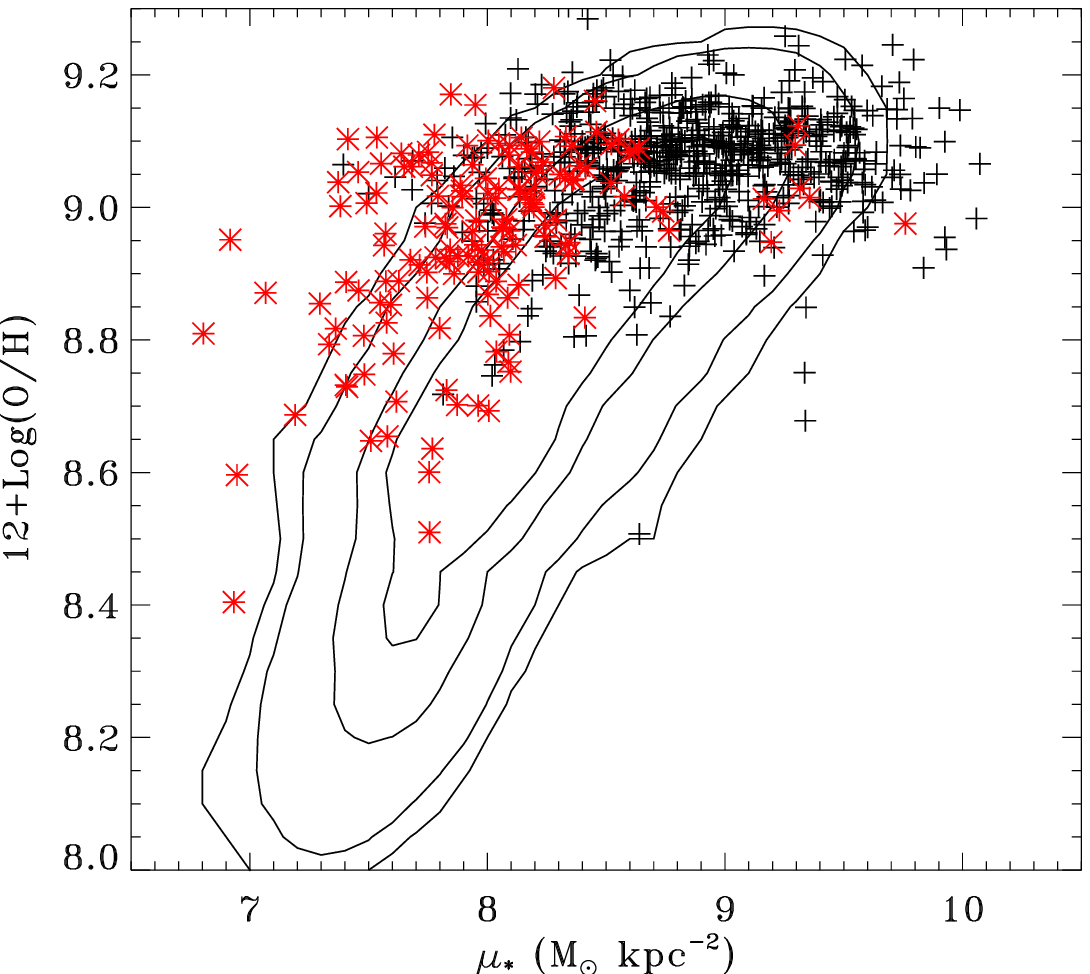}
\includegraphics[width=\columnwidth]{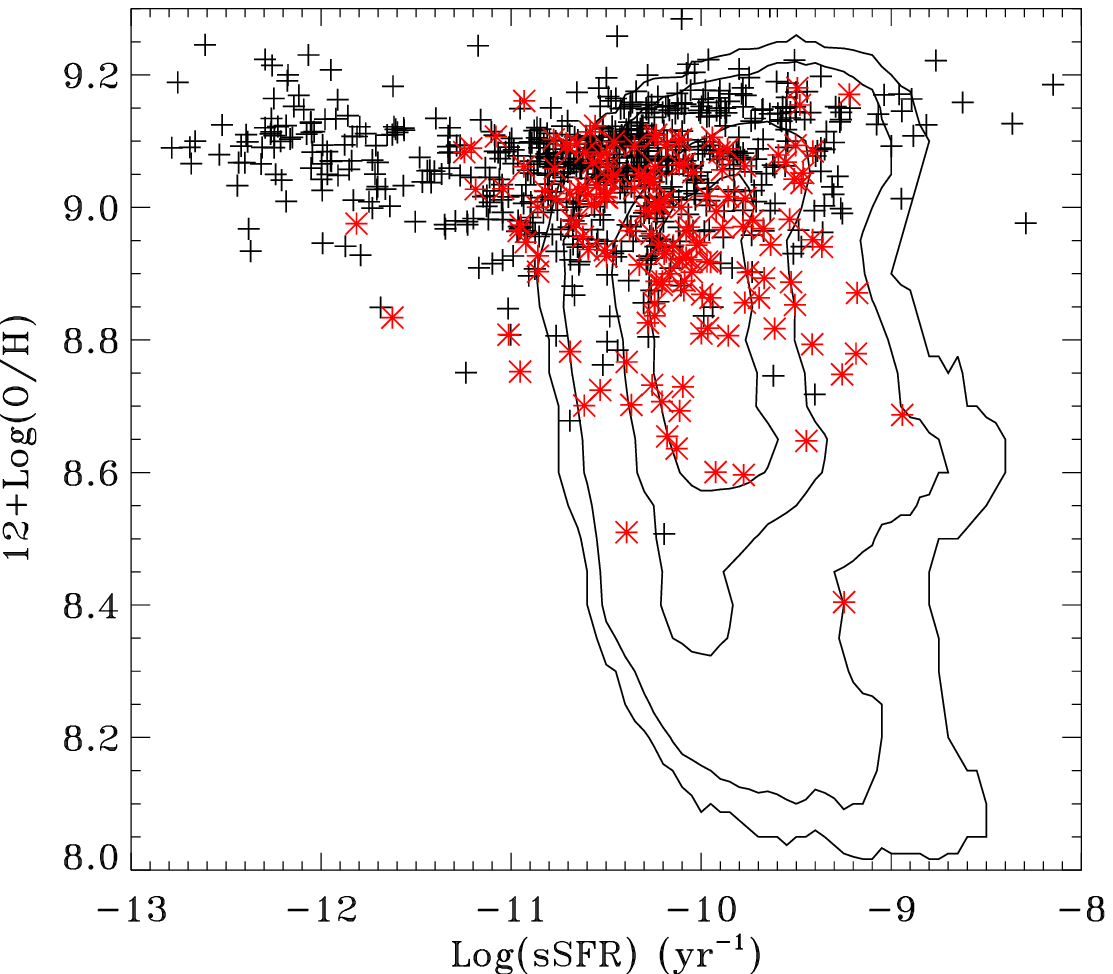}
\caption{\label{inner_outer} {\bf Left:}
  Metallicity versus projected stellar mass surface density ($\mu_*$),
  for all independent spatial bins within 0.7$R_{90}$ (black crosses) 
  and outside 0.7$R_{90}$ (red asterisks). Top panel shows O3N2
  metallicities, while bottom panel shows T04 metallicities. Contours show where 33\%, 67\%,
95\% and 99\% of SDSS galaxies in the same redshift range lie (including all stellar masses), 
where $\mu_*$ and metallicity are for the region directly under the central 3\arcsec fiber.  
The $\mu_*$--metallicity relation for SDSS galaxy centers is similar to that for
GASS galaxies, but exhibits a slightly different downturn.
{\bf Right:} Metallicity
  versus dust-corrected specific star formation rate (sSFR), with
  symbols and contours coded as before, and again with O3N2
  metallicity in top panel and T04 in bottom. Low metallicity regions in galaxy outskirts
exhibit increasing sSFR with decreasing metallicity. Lack of contours 
at low sSFR in the bottom panel is a selection artifact, as T04-style 
metallicities are not reported in the MPA/JHU catalogs for such low star
formation rates.
}
\end{figure*}

In Figure~\ref{sings_compare}, the converted SINGS points are plotted
versus radius, as open diamonds. We include only points from 
the 15 SINGS galaxies with $Log(M_*)>10$ and $>8$ measured points, 
and for comparison we plot our GASS points from 
Figure~\ref{metal_r90} as small circles. First, we note that the SINGS
points from these high-mass galaxies are in quite good agreement with
our own across most of the radial range probed, including the drops we
observe at high radius. However, SINGS data in some cases appear to reach higher
metallicities at the centers of galaxies, compared to our own.

Since GASS probes more distant galaxies, features on spatial scales of $<0.1
R/R_{90}$ begin to fall below the resolution of the data, and one
might worry that we bin over too large an area to detect these elevated
central metallicities. This could contribute to the overall flatness
of the gradients we see. However, SINGS galaxies exhibit
high-metallicity at radii as large as $\sim0.3$~$R/R_{90}$, 
beyond the regime where limited spatial resolution could be affecting the GASS
measurements. 

Similarly, one might worry that the difference arises because 
\citet{moustakas10} compiled measurements of isolated HII regions, 
while our data are integrated over a broader area, and so may 
include a component from the diffuse ionized medium associated with 
older star formation or shock-ionized gas \citep[e.g.,][]{dopita06}.
However, within the regime spanned by our data, \citet{dopita06}
showed that the bias between measurements of HII regions and whole
galaxies is small, and we calculate that it would lead to a metallicity
offset of $<0.03$~dex, smaller than the offset observed, and
less than half our quoted uncertainty.

We suspect, instead, that the slightly higher SINGS points
are either an artifact of the conversion formula used
to place SINGS data on the O3N2 scale, or else a reflection of
the inherent limits to uniformity in any literature compilation 
like \citet{moustakas10}. First, we note that these
high metallicity points ($12+Log(O/H)>8.95$) lie outside of the full 
range of metallicities used by \citet{kewley08} in deriving 
their conversion formula, and so the conversion
from the original R23 to our O3N2 for these points is at least a 
mild extrapolation. Furthermore, the highest SINGS points
 at $12+Log(O/H)>8.95$ belong almost entirely to two galaxies, 
NGC3351 and NGC5194 (M51), hinting that systematic offsets 
between literature sources could be important.

In general, however, our new
results on the large, unbiased GASS sample, are in good agreement with
previous work, but fill in the picture with a more 
complete sampling of the full radial range. Though a more robust 
investigation of how metallicity gradients vary with mass must await 
more data extending to lower masses, we have shown in this Section 
that galaxies at high masses have predominantly flat inner metallicity 
profiles, and that this result is not in conflict with earlier work 
once we compare within the same stellar mass range.

\subsection{Outer Metallicity Drops and the $\mu_*$-Metallicity Relation}
Having established that inner metallicity profiles are largely
flat except for those near the low end of our mass range, we
wish to understand why some galaxies appear to show precipitous
drops in metallicity at or near $R_{90}$. First, we note that galaxies with these
steep metallicity drops occur in galaxies of  all stellar masses: 
low metallicity points can be
seen at large radius in all four stellar mass bins plotted in
Figure~\ref{metal_by_mass}. In asking what might be driving these
drops, we must first characterize them in two ways: 
first, what types of galaxies do these low-metallicity regions reside 
in, and, second, what are the internal, local conditions
like at each of these low-metallicity sites? In this section,  
we will examine the link between outer metallicity and 
{\it local}  properties within the galactic disk, and defer the
question of global galaxy properties to \S5.

In Figure~\ref{inner_outer}, we plot
metallicity as a function of local         
stellar mass density ($\mu_*$, left panels) and specific star formation rate
(sSFR, right panels). Results are shown for both the O3N2 (top) and T04 (bottom)
metallicity calibrations. Interestingly, the $\mu_*$--metallicity
relation evaluated {\em within galaxies}  
bears striking resemblance to the well-known global mass-metallicity
relation (e.g., T04). There is a relatively tight
correlation between metallicity and local surface density at
$\mu_*$ less than $3 \times 10^8 M_{\odot}$ kpc$^{-2}$ 
($|\rho|=0.38$, $8.5\sigma$), followed by a ``turnover''
where the dependence of metallicity on stellar surface mass density
weakens, similar to
the way the global mass--metallicity relation flattens at high masses.

In Figure~\ref{inner_outer}, data marked with red asterisks are from regions with 
$R>0.7R_{90}$. It is clear, therefore, that
low metallicity regions in the outskirts of galaxies are generally
associated with {\it low} underlying stellar mass density.
Likewise, in the right-hand panel of Figure~\ref{inner_outer}, one can see that
in the outer disks
the lowest metallicity regions tend to be those with the highest
ratios of current star formation to pre-existing stellar mass (i.e., sSFR). 

As in the $\mu_*$-metallicity relation, points closer to the centers 
of galaxies do not show a significant correlation between sSFR and
metallicity. Of course, this was expected given that inner
metallicities are nearly invariant, as discussed above, but it is 
still interesting to
note that sSFR in these inner regions spans four orders of magnitude
without appreciably affecting metallicity.

The contours overplotted in Figure~\ref{inner_outer} show where SDSS 
galaxies in the GASS
redshift range (but spanning all stellar masses) lie on these relations. Recall that for these
galaxies,   $\mu_*$, metallicity, and sSFR pertain to the central 
regions of the galaxies falling directly under the SDSS fiber. 
We find that the central regions of SDSS galaxies
also follow a similar  $\mu_*$--metallicity relation, but there
is a somewhat sharper decline in metallicity with decreasing
$\mu_*$, or perhaps a shift in the threshold value of $\mu_*$ where
metallicity begins to decline, compared to the outskirts of GASS
galaxies.

Since the low-metallicity points in the
SDSS sample come from the centers of galaxies with much lower stellar
masses than the GASS sample, it may not be surprising that the two
relations differ slightly. Specifically, one standard 
explanation for the mass-metallicity relation says that
low mass galaxies exhibit lower metallicity because a higher fraction
of their metals are driven by supernovae winds out of the galaxy
due to their shallower potential wells (T04; Dalcanton 2007). Since even
the outskirts of our massive GASS galaxies have higher escape
velocities than the centers of low-mass ($\sim10^9 M_\odot$) galaxies, 
such an effect could be
responsible for the downward offset of these galaxies' centers relative 
to the relation that we see.
Other explanations are also possible, including a secondary dependence
on sSFR similar to that advocated by \citet{mannucci10}, but a 
fuller exploration is beyond the scope of this paper. 

Regardless of the source of the offset, 
the fact remains that our outer disk resolved points occur in 
very different locations within galaxies of very different masses than
the SDSS sample, but show a broadly similar relation between
metallicity, stellar mass density, and specific star formation rate. 
One possible explanation is that the correlation between metallicity 
and local stellar surface density is more fundamentally a correlation 
between metallicity and local gas mass fraction ($\mu_g/\mu_*$). In regions of 
low $\mu_*$, gas mass fractions are likely to be high (either
at the centers of low-mass SDSS galaxies or the edges of high-mass GASS 
spirals), and the lower metallicities could be reflecting a less complete
processing of the gas into stars.
Since one of the key strengths of the GASS sample is that we have
direct measurements of the gas content, we are in a unique position to
test whether sub-solar metallicities and elevated sSFRs at our
galaxies' outskirts are linked to their reservoirs of gas.

\begin{figure*}
\includegraphics[width=\columnwidth]{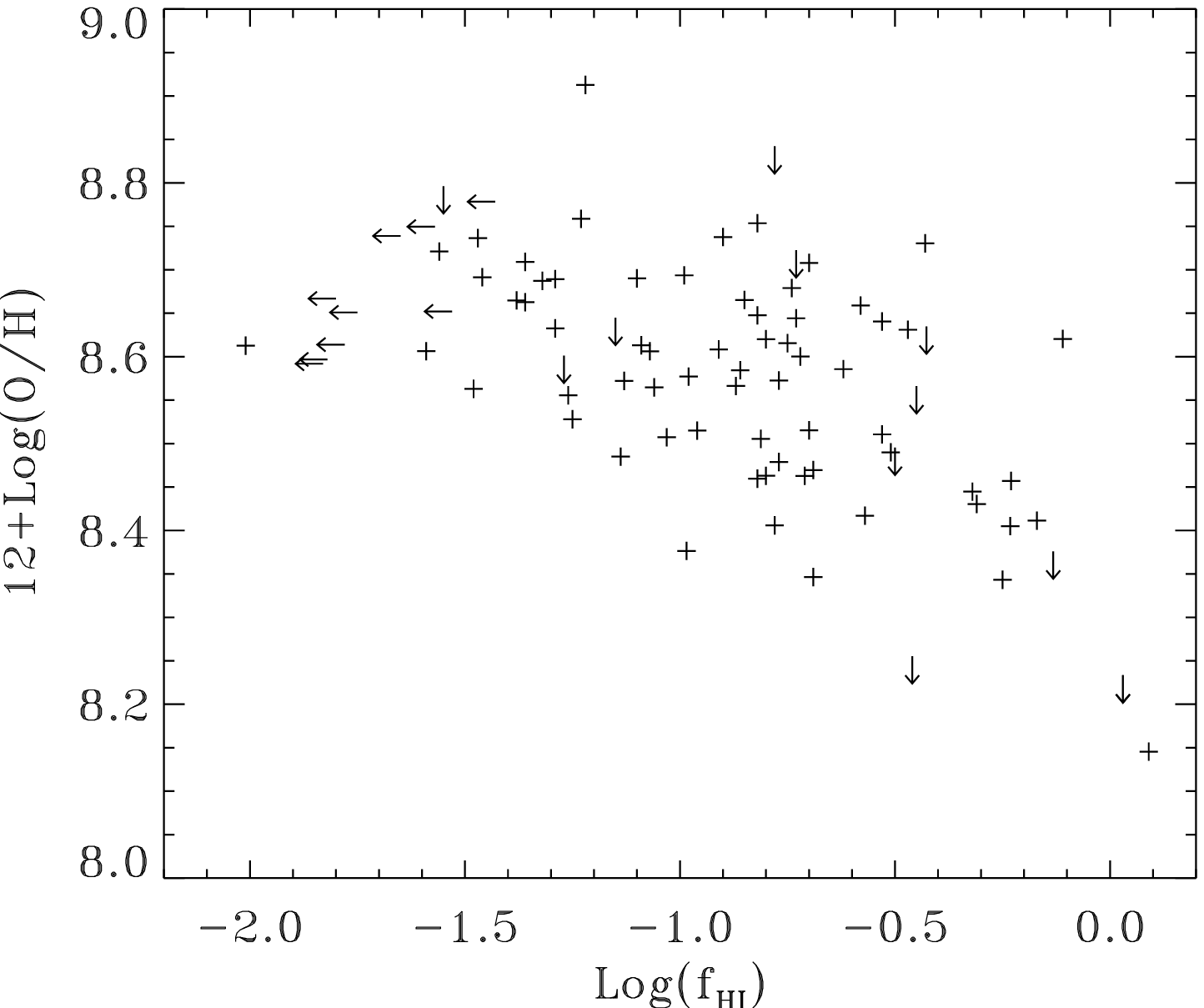}
\includegraphics[width=\columnwidth]{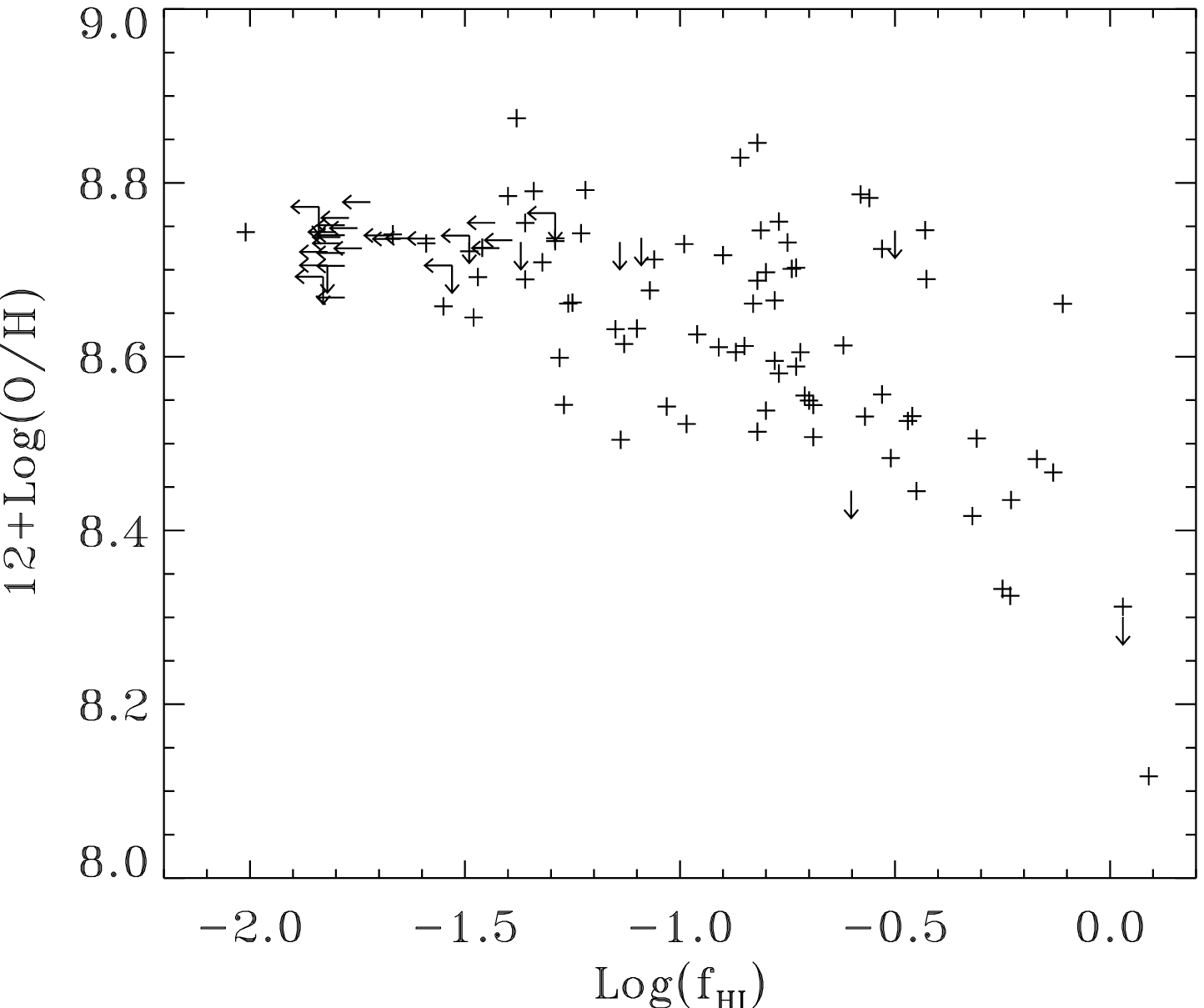}
\caption{\label{FP} Galaxy outer disk metallicity as a function of total galaxy HI 
fraction, $f_{HI}$. Left-hand panel displays the lowest-measured metallicity 
point for each galaxy, while the right hand panel shows metallicities
measured from the integrated spectrum of all flux from $R>R_{90}$ for each galaxy.
Both show that outer-disk metallicity is well-correlated with total HI content.}
\end{figure*}

\section{Discussion}
\subsection{The Relation Between Gas Content and Metallicity}

\begin{figure*}
\includegraphics[width=0.66\columnwidth]{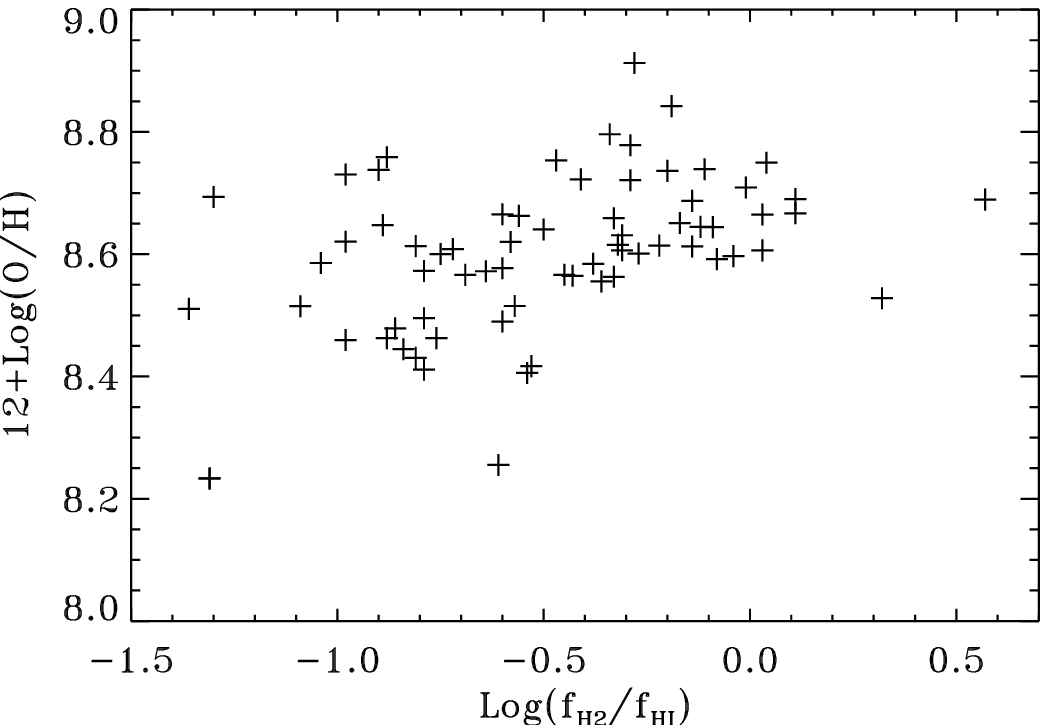}
\includegraphics[width=0.66\columnwidth]{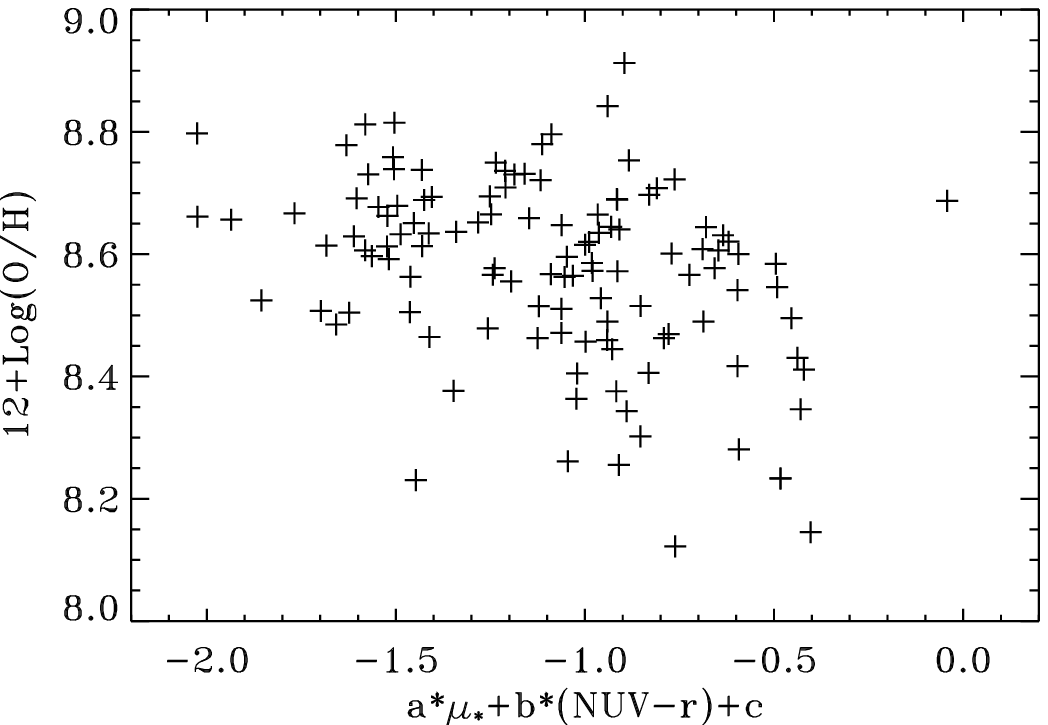}
\includegraphics[width=0.66\columnwidth]{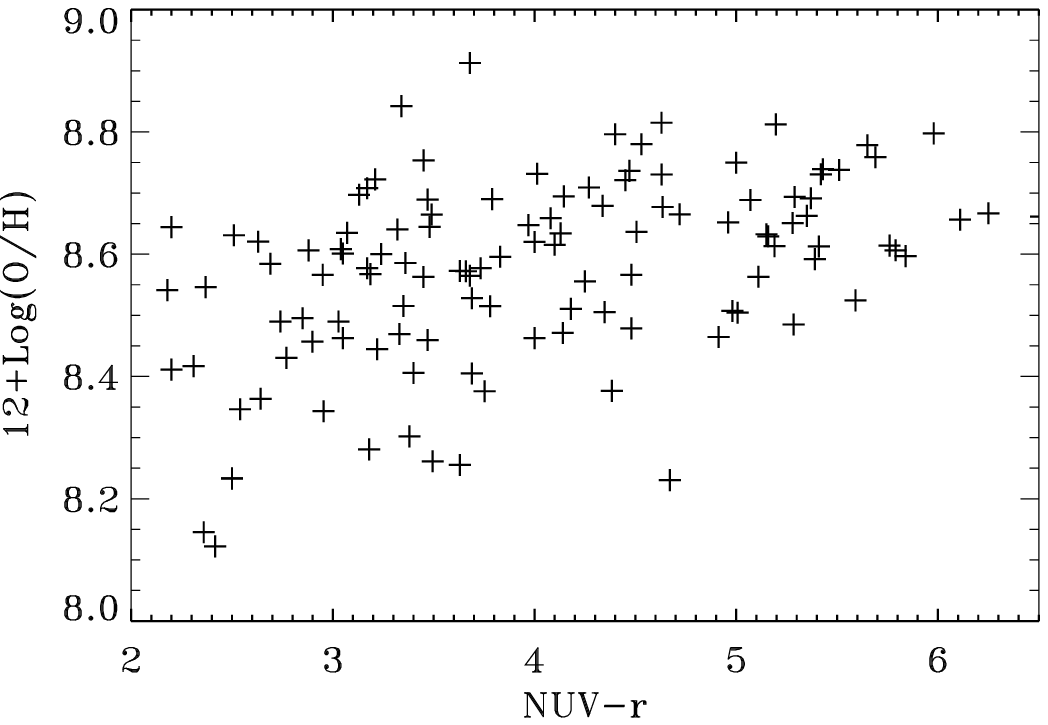}
\\
\includegraphics[width=0.66\columnwidth]{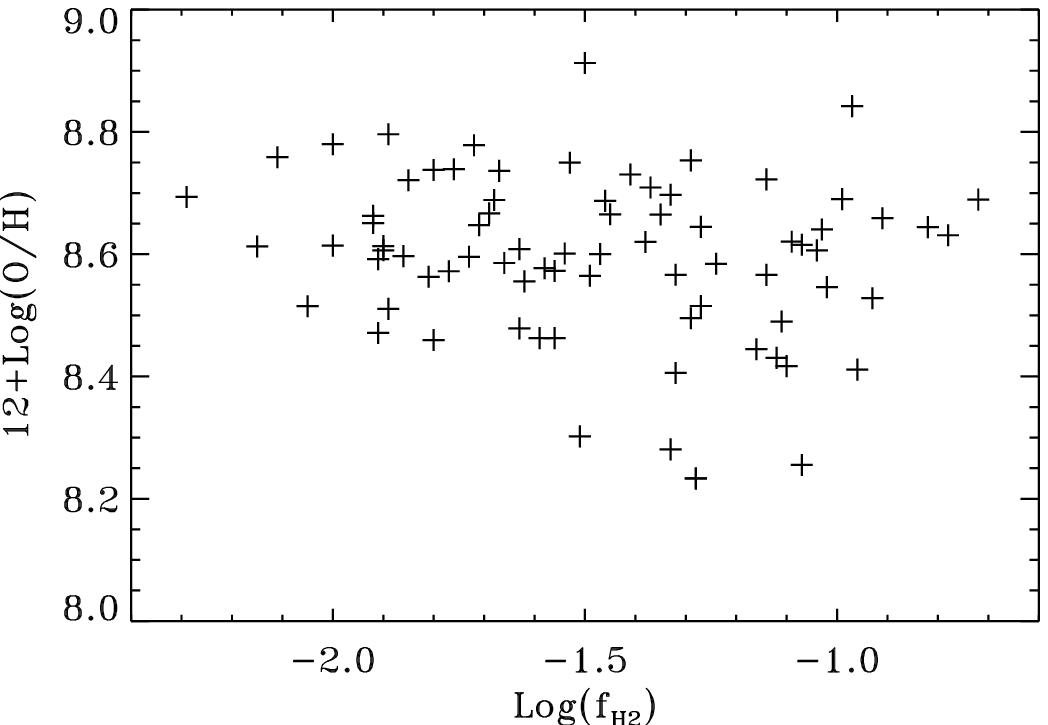}
\includegraphics[width=0.66\columnwidth]{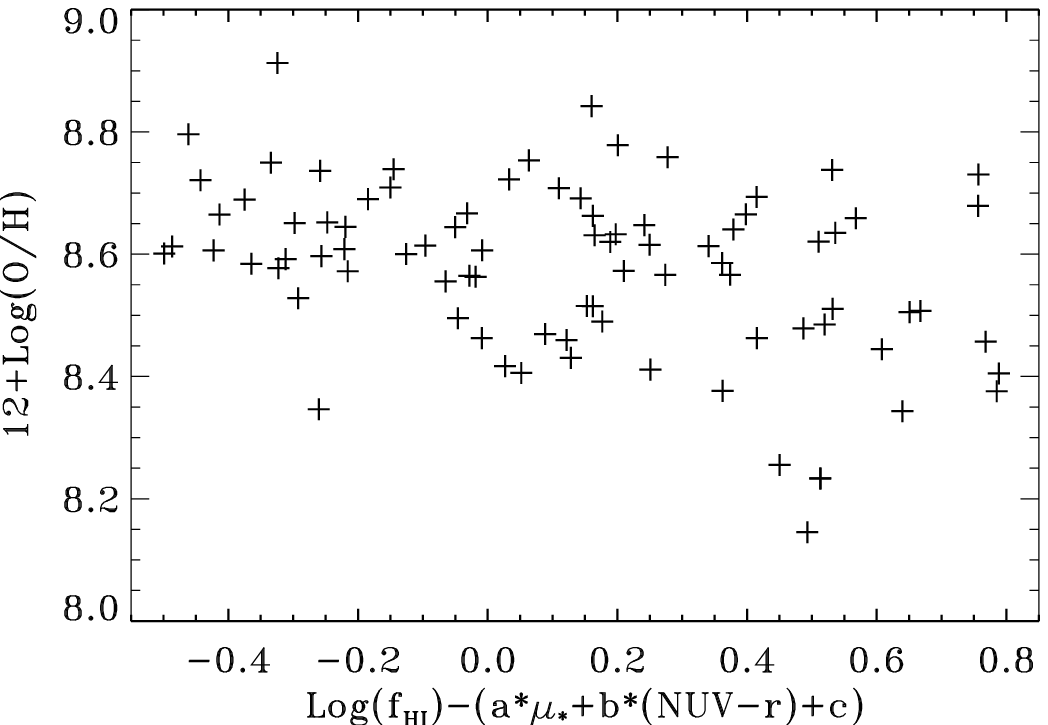}
\includegraphics[width=0.66\columnwidth]{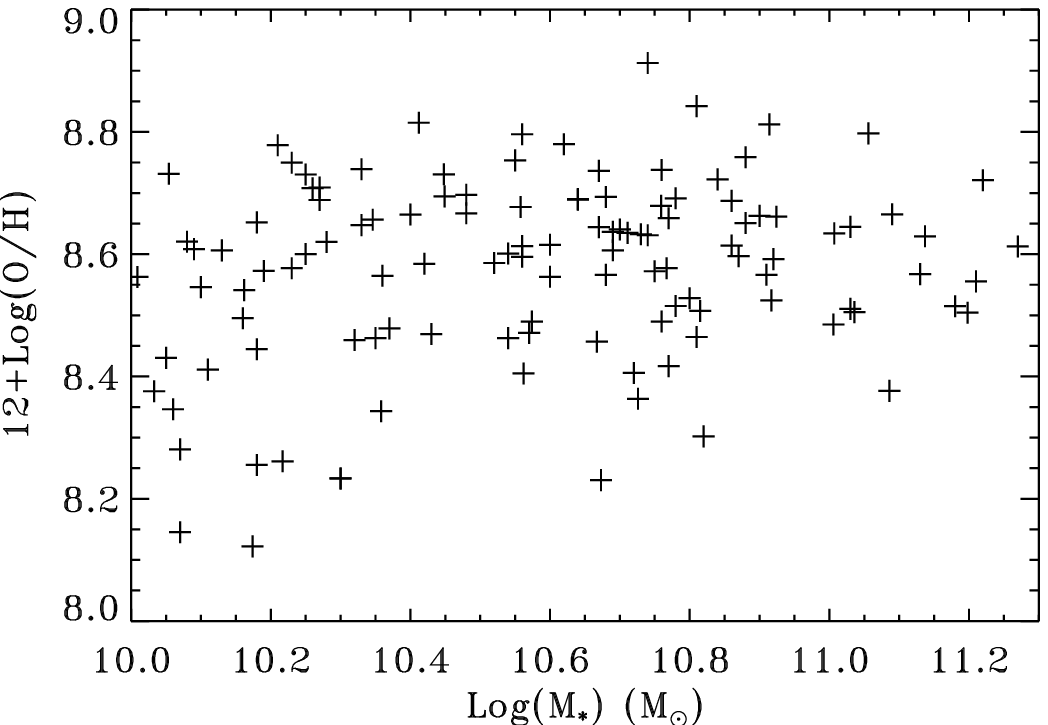}

\caption{\label{global_plots} Outer metallicity, calculated as in the left panel of Figure~8, 
plotted as a function of, from top left to bottom right, ratio of molecular to 
atomic gas ($f_{H2}/f_{HI}$), the combination of NUV--{\it r} and full-galaxy $\mu_*$ 
found to best predict HI content by C10, NUV--{\it r} color alone,
molecular gas fraction ($f_{H2}$), residuals from the C10 HI plane, 
and stellar mass ($M_*$). 
None predict outer metallicity as well as the HI fraction ($f_{HI}$) alone.}
\end{figure*}

In C10, we found that the HI content of a given
GASS galaxy was most accurately predicted by an fundamental-plane style
``HI plane'' constructed from whole-galaxy measurements 
of $\mu_*$ and NUV$-r$ color 
(which is a good proxy for sSFR). Since we have seen in the above
section that local $\mu_*$ and sSFR are also tied closely to the
local gas-phase metallicity, it is natural to wonder if we can identify
a direct link between metallicity and gas content.

Remarkably, as can be seen in Figure~\ref{FP}, we have found that there is a
surprisingly tight relation between the global HI fraction
($f_{HI}=M_{HI}/M_*$) and the lowest metallicity point that we measure
($R>0.7R_{90}$, O3N2 scale) for each galaxy. We measure a correlation 
coefficient $|\rho|=0.53$, significant at the $5\sigma$ level. 
We include upper limits as if they were detections for this calculation, but
excluding these limits---either in $f_{HI}$, metallicity, or both---does not
significantly change the result.

The C10 HI plane can be used to predict $f_{HI}$ with rms
accuracy of $\sim0.3$~dex, and for comparison the scatter in our outer
metallicity-$f_{HI}$ relation is 0.4~dex in $f_{HI}$. 
We note, however, that outer metallicities that are near solar provide 
little predictive power for HI. In contrast, low metallicity points
{\it do} seem to be universally associated with gas-rich galaxies. 
Whether or not Figure~\ref{FP} reflects a {\it continuous} relation
between outer metallicity and $f_{HI}$, or whether it is more
properly described as a {\it threshold} in $f_{HI}$,
above which outer metallicities are suppressed in proportion to the gas
content, is unclear from the
current data. We hope to revisit this issue with the full GASS data-set
once it is in hand.

The quantity plotted in the left panel of
Figure~\ref{FP}, being defined
as the lowest metallicity we measure in a galaxy's outskirts, could be
subject to a number of biases due to the imprecise definition.
To check whether the way we define the
metallicity value is driving this relation in any way, we also
calculated metallicity values by coadding all spectroscopic flux
from $R>R_{90}$ into a single high-S/N spectrum for each galaxy. 
Metallicities measured from these spectra, being integrated over a
much larger area of the galaxy, often do not reach such low
values as our more finely sampled points. Even so, when
plotting these metallicity values against $f_{HI}$, in Figure~\ref{FP}, right panel,
we find a nearly identical relation as before, with a marginally higher
$|\rho|=0.57$ and $6\sigma$ significance. 
We note, also, that plotting the size of the
outer metallicity {\it drop}, rather than the outer metallicity directly, gives
largely the same results.
Considering that the inner metallicities of most GASS galaxies
are nearly solar everywhere, this is not surprising.

To understand what might be driving this peculiar relation between
a very {\it local} quantity, the outer metallicity, and the {\it
  global} gas content, we must examine how outer metallicity relates
to other global galaxy characteristics. In Figure~\ref{global_plots},
we plot the minimum outer metallicity on the O3N2 scale 
(as in the left panel of Figure~\ref{FP}), as a function of several 
such quantities.

The first quantity of interest is the molecular gas fraction,
$f_{H2}$. As can be seen in the lower-left panel, $f_{H2}$ by itself
does not correlate with outer metallicity. Since H$_2$ is generally
more centrally concentrated in galaxies than HI \citep[e.g.,][]{leroy09}, this result may
not be surprising. In fact, we do detect a weak correlation 
($|\rho|=0.39$, $3.9\sigma$) between $f_{H2}$ and {\it central} metallicity. 
The outer metallicity may simply be more closely 
associated with the size of a galaxy's gas {\it reservoir}, as traced 
by HI, rather than by the amount of currently star-forming gas, 
traced by H$_2$.

Interestingly, the {\it ratio} of molecular to
atomic gas, shown in the top left panel, {\it does} correlate with
outer metallicity, in the sense that galaxies more dominated by their
HI exhibit lower outer metallicity. However, this correlation is
weaker ($|\rho|=0.44$) and less statistically significant ($3.7\sigma$)
than the HI relation in Figure~\ref{FP}, and so we may only be seeing
here a reflection of the underlying relation between HI and outer metallicity.

\begin{figure*}[t]
\includegraphics[width=\columnwidth]{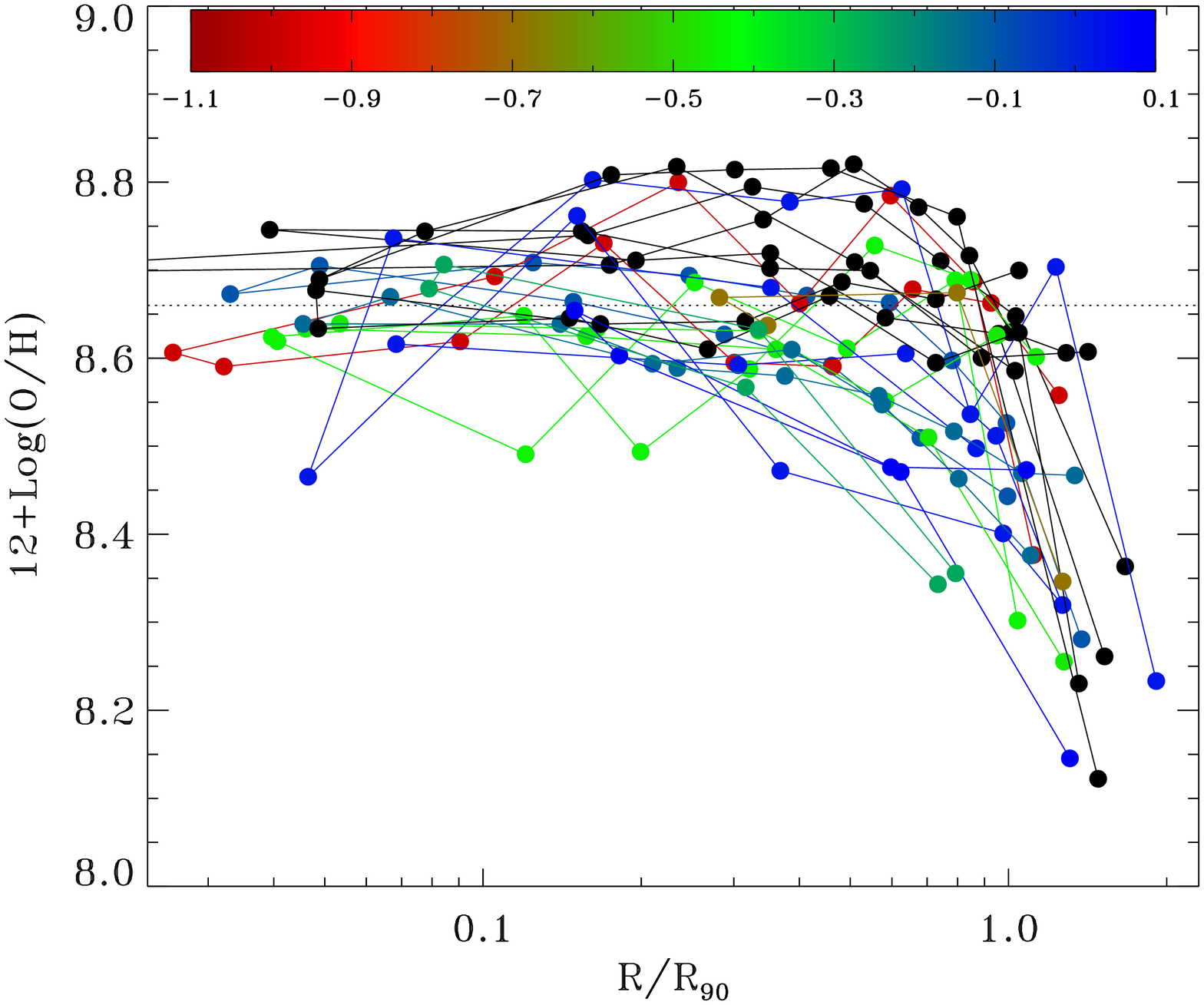}
\includegraphics[width=\columnwidth]{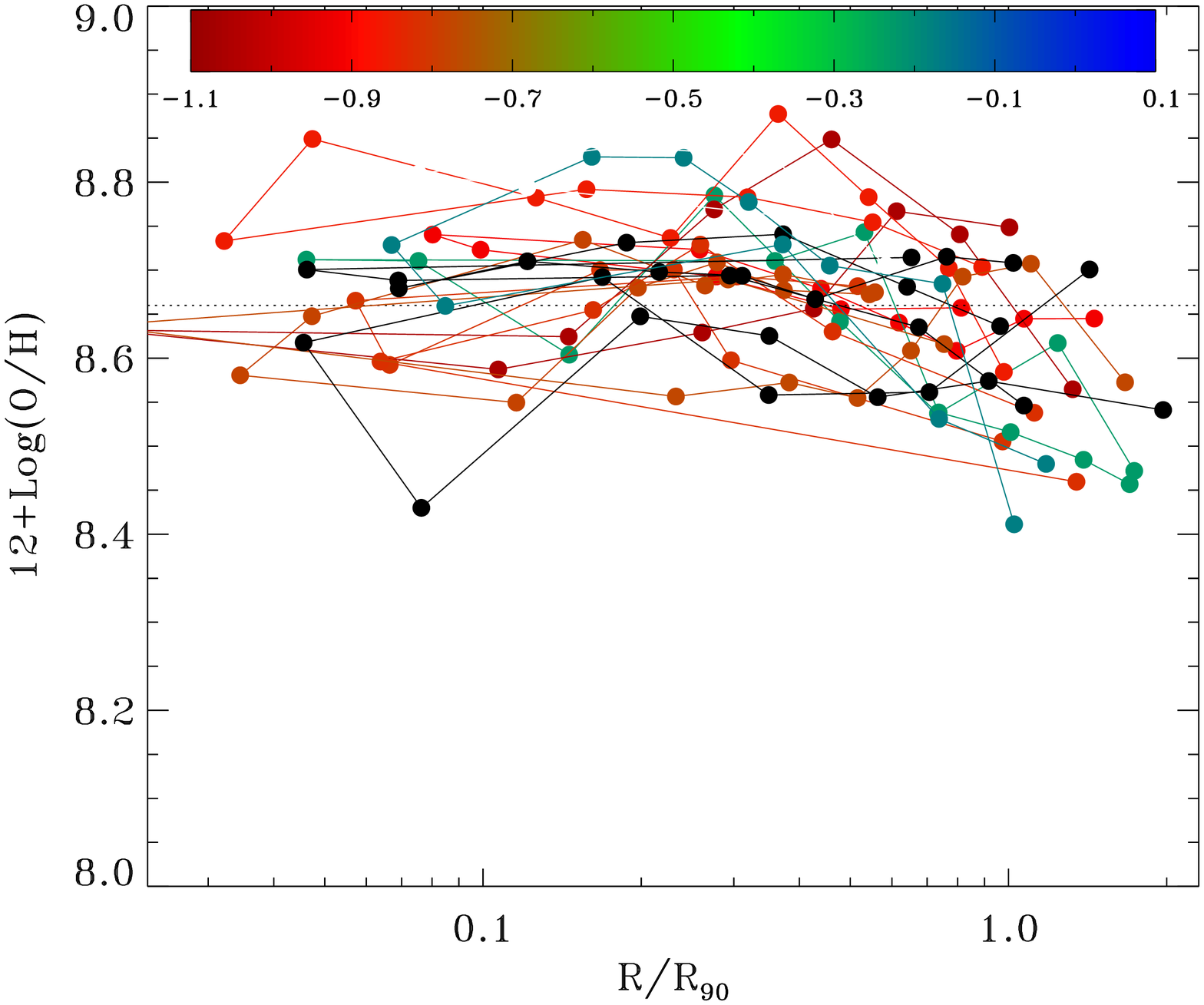}
\caption{\label{metal_gradients} Left: Metallicity versus $R/R_{90}$ for the
13 galaxies with the largest outer metal drops. Right: Metallicity versus $R/R_{90}$ for the
matched comparison galaxies.
In both panels, solid lines connect adjacent measurements
for each individual galaxy, and both halves of each galaxy are plotted separately, `folded'
onto the same $R/R_{90}$ scale. Points and lines
are color coded for galaxy HI content, according to the color bar at top, which indicates 
$Log(f_{HI})$ associated with each shade. Galaxies that do not yet have an HI measurement
are plotted in black.}
\end{figure*}

In the upper middle panel of Figure~\ref{global_plots}, we plot outer metallicity as a function
of the same combination of global NUV--{\it r} and $\mu_*$ that goes into the C10
HI plane--i.e., the x-axis can be thought of as the $f_{HI}$
that is predicted given the galaxy's NUV--{\it r} and $\mu_*$. 
We do this to evaluate the possibility that the
$f_{HI}$--metallicity relation is simply an induced correlation: if it is really
the star formation rates and stellar mass densities that set metallicity, as our results above seem 
to suggest is true locally, it may be that the correlation with HI content is not fundamental,
but instead driven by the dependence of HI on the same underlying parameters
(sSFR and $\mu_*$, but galaxy-averaged now) that drive metallicity.

What the upper middle panel
of Figure~\ref{global_plots} shows, however, is that the correlation
of outer metallicity with global $\mu_*$ and NUV--{\it r} is 
relatively weak compared to Figure~\ref{FP}, with $|\rho|=0.37$ and a 
significance of $4\sigma$. Since C10 found the 
relation between $f_{HI}$ and NUV--{\it r} color alone was nearly as 
tight as the one including $\mu_*$, we also plot in the upper right
panel outer metallicity versus NUV--{\it r}. The correlation here is 
more significant, at 4.8$\sigma$, but again the correlation 
coefficient is not quite as strong ($|\rho|=0.44$) as we saw in Figure~\ref{FP}.

The {\it residuals} from the C10 HI plane
are useful for identifying galaxies that have either abnormally high
or low atomic gas content, compared to that expected for galaxies with
their specific combination of $\mu_*$ and NUV--{\it r}. In the lower
middle panel of Figure~\ref{global_plots}, we plot these residuals
versus outer metallicity. The correlation is quite low and only
marginally significant ($2.8\sigma$), suggesting that it is the
absolute level of HI content that matters, rather than the {\it
  excess} or {\it deficit} of HI compared to other similar galaxies.

Finally, we checked for relations between metallicity
and other global galaxy properties such as concentration,
$\mu_*$, and $M_*$. None show any significant correlation; the lower
right panel of Figure~\ref{global_plots} shows metallicity versus
Log$(M_*)$, which is representative of the others. Outer metallicity
thus appears to depend hardly at all on the structural properties of
galaxies in GASS.

\begin{figure*}
\centering
\includegraphics[width=2\columnwidth]{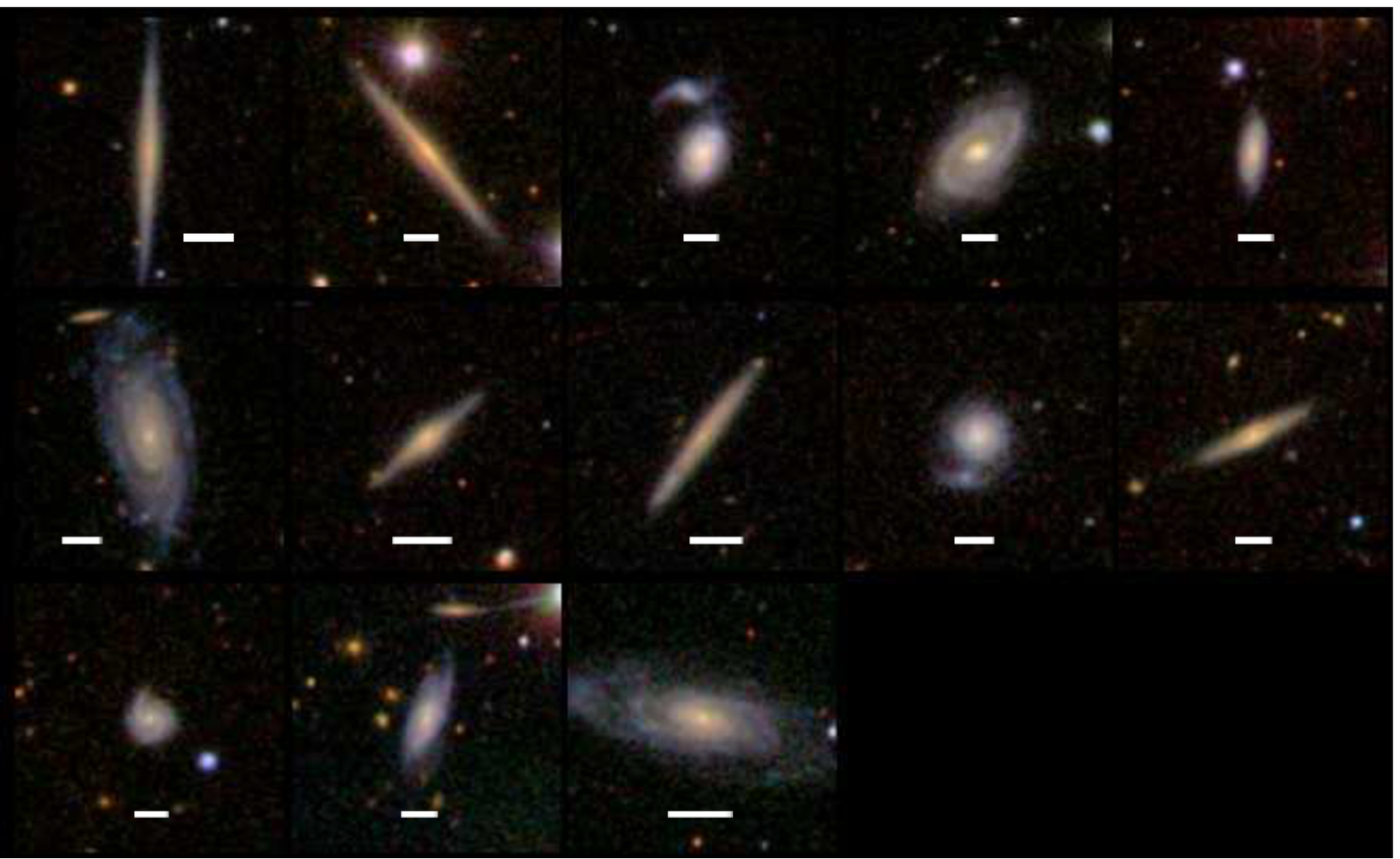}
\medskip

\includegraphics[width=2\columnwidth]{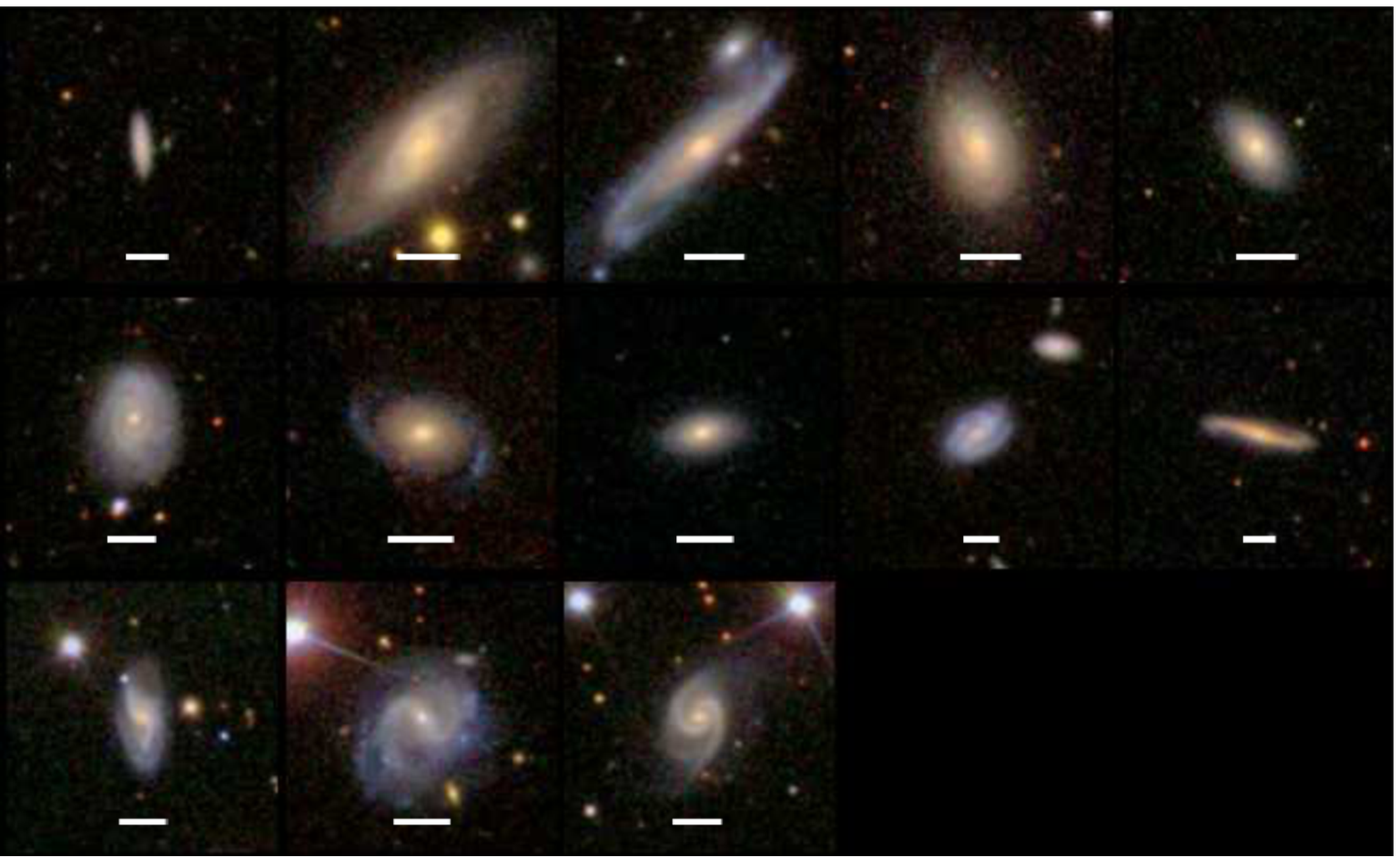}
\caption{\label{low_stamps}SDSS postage stamp images of galaxies with the steepest outer
metallicity drops (top), and the matched control sample selected
without regard for metallicity (bottom). 
Images are $80\arcsec\times80\arcsec$, and are arranged so that the
match to a galaxy in the top panel is in the same relative position
in the lower panel. White bars in each image indicate a projected physical length of 10kpc 
at the redshift of that galaxy.}
\end{figure*}

In short, we have found that the relation between $f_{HI}$, a global
property of the galaxy, and outer metallicity $12+Log(O/H)$, a very
local one, is tighter and therefore likely {\it more} fundamental 
than any relation between global SFR/$\mu_*$ and metallicity. 
The question, then, is why this should be true.
One possibility is that metallicity at the outer edge of the star-forming disk
is simply a sensitive `thermometer' of sorts for measuring the amount of new
gas accreting onto the disk of the galaxy, or perhaps for the rate
that existing gas is transported inward. Lower metallicities would
simply be reflecting a higher proportion of pristine or relatively
unenriched gas residing in or flowing through the outer stellar disk. 

If this is correct, then varying rates of gas accretion/flow may leave
other signatures in, for example, the stellar populations and star 
formation rates of galaxies, or on their radial gradients.
To evaluate this possibility, in the following section we will examine in 
detail the subset of galaxies exhibiting the strongest
metallicity drops in our sample, and compare to a control sample
selected without regard to outer metallicity.

\subsection{The Galaxies with Steepest Metallicity Drops}

In order to determine whether a low outer metallicity is associated with
any distinctive features in the star formation rates and histories of 
our galaxies, either at their outer edges or across their 
disks, we select for detailed study a small sample containing
only those galaxies with the steepest drops in outer metallicity.
We will refer to this as the  `low metallicity' or `large drop' sample, 
and we will compare it to a `control' sample selected without
regard to metallicity, but matched in global characteristics one-for-one
with galaxies in the large-drop sample.
We note that both samples are selected from the subset of 119 galaxies that exhibit
measurable star formation in their outskirts at $R>0.7R_{90}$, which
ensures that all the galaxies in the control sample have significant star
formation and measurable metallicities at the same large radii as the
large-drop galaxies.

In total, we identify 13 galaxies with outer-disk metallicities
$12+Log(O/H)<8.4$, a threshold we chose both because no measured point
in the inner region of any galaxy reaches this low, and because it is
significantly below solar at the $>3\sigma$ level (0.25~dex). 
Such galaxies make up about 10\% of the total sample, and so even 
though the number
identified is small, they represent a significant
proportion of all massive star-forming galaxies. 
For the control sample, we select each matched counterpart by
requiring that it be within 0.2~dex in stellar mass, 0.4~dex in global
$\mu_*$, and 0.3 in NUV-{\it r} color, similar to the
procedure in \citet{wang11a}. 
These limits were chosen to ensure that
each large-drop galaxy has at least one match within our spectroscopic
sample, and for cases with more than one match we select randomly
among the choices.

In Figure~\ref{metal_gradients}, in the left-hand panel we plot the radial metallicity
profiles of all 13 of our low-metallicity galaxies, as a function of $R/R_{90}$.
Solid lines connect adjacent points to more easily follow the two
folded halves of each profile (i.e., from both sides of each
galaxy). Points are color coded for galaxy HI content, as illustrated
by the color bar.
It is clear from this plot that even those galaxies with the
strongest metallicity drops show typically flat profiles in their
inner regions (though some variation can, indeed, be seen). The median
difference between the inner metallicity and the lowest measured point
is $\sim0.3$ dex for these objects. In the
right-hand panel of Figure~\ref{metal_gradients}, we now plot the metallicity profiles of
the control sample. Again, we see quite flat inner profiles, though
there may be a hint that even these galaxies show subtle drops in
metallicity at the highest measured points. 
Consistent with the overall correlation shown in Figure~\ref{FP}, galaxies
with large metallicity drops have on average higher $f_{HI}$, which
can be seen by noting the quite different range of colors between
the two panels.

\subsubsection{Morphologies}
Our low-metallicity and control samples of galaxies by design have certain
features in common, in particular their stellar masses and colors, 
as well as widespread star-formation extending out to $R>R_{90}$.
Yet the disparity in the magnitude of the metallicity drop, as
well as the wide variation in HI content, suggests that it is
worthwhile to examine the images of both samples directly, to check
for any subtle structural differences between the two that are not
apparent when considering just the broadest characteristics.

In Figure~\ref{low_stamps}, we show SDSS postage stamp images of these
galaxies, with large-drop galaxies in the top panel, and the control
galaxies at bottom. Images are arranged so that each matched pair of
galaxies is in the same relative location in the top panel and bottom.
We remind the reader that galaxy pairs were not matched in redshift (beyond
the normal GASS constraint of $0.025<z<0.05$), so
differences in apparent size are to be expected. For aid in comparison,
white bars beneath each image indicate 10kpc projected distance at the redshift 
of each galaxy.

The first striking feature of Figure~\ref{low_stamps} is the large number of
relatively edge-on galaxies in the low-metallicity sample. There are
two possible explanations for this trend, both related to observing
geometry. First, if the lowest
metallicity regions are distributed unevenly, in a stochastic manner
around the outer edges of galaxy disks, then an edge-on galaxy may
afford us a higher probability of actually observing one, as the light
collected by our slit has passed through a larger proportion of the galaxy.
As can be seen in the Appendix, where individual profiles of 
large-drop galaxies are shown, some of these galaxies (but not all) 
show a metallicity drop on one side only, which might argue in favor 
of this scenario.

Alternatively, surface brightness enhancement due to viewing the
galaxy edge-on could simply enable our fixed-integration time observations to
reach further out in an edge-on galaxy, where such low-metallicity
star-forming regions are perhaps more likely to reside. We do note
that the maximum radius reached for edge-on galaxies is somewhat
higher than for other objects (median of $1.2R_{90}$ for $b/a>0.5$ vs
$1.4R_{90}$ for $b/a<0.5$), but it is unclear if this difference is
large enough to cause the observed bias. The median radius reached for the control
sample galaxies is only slightly lower than the large drop galaxies: $1.2R_{90}$ 
and $1.3R_{90}$, respectively.

A few galaxies in Figure~\ref{low_stamps} seem to harbor
close-in tidally disturbed companions, which could suggest that new gas
accreted or cannibalized from these companions is the source of the
low-metallicity material. However, the majority of the large-drop
galaxies do not have obvious companions, and the frequency of
companions is not obviously larger than that seen in the control
sample images. 
Indeed, the one large-drop galaxy that we have previously
studied in detail \citep{moran10}, shows remarkably little
evidence for any sort of dynamical disturbance, such as one would
expect in the case of an accreted companion. Though beyond the scope
of this paper, rotation curves are available for all of these
galaxies, and we expect to address the relation between gas content and
galaxy dynamics and/or mergers in a future paper.

We measured a number of quantitative morphological parameters for both sets of galaxies
to search for any differences. These included asymmetry, bar fraction, and an 
asymmetry variant weighted to the outer disk \citep{wang11b}.
Though such small samples make it difficult to assess any statistical difference between
the two sets of galaxies, we find no clear evidence that the populations are different
in any of the key morphological measures. This remains true whether or not we exclude the
most edge-on galaxies (which could have peculiar values of these morphological parameters).

Besides the obvious difference in orientation, then, we find no dramatic
differences in appearance between the large-drop and control sample
galaxies in Figure~\ref{low_stamps}. 
Below, we will examine the radial profiles of quantities
detailing the star formation histories and current star formation
rates as a function of radius across each galaxy type.

\subsubsection{Star Formation Histories}

So far, then, our large-drop sample of galaxies appears to differ from
the control sample only in two important properties:
metallicity drop and HI content.
But the key question remains: why does a high HI content appear to drive 
low-metallicity star formation in galaxies' outskirts?
To help answer this question, we can look to the star formation rates and
histories of our two groups of galaxies. Specifically, we 
can examine the radial profiles of a number of spectroscopic
diagnostics of both past-averaged and present-day star formation.

\begin{figure*}[t]
\includegraphics[width=0.66\columnwidth]{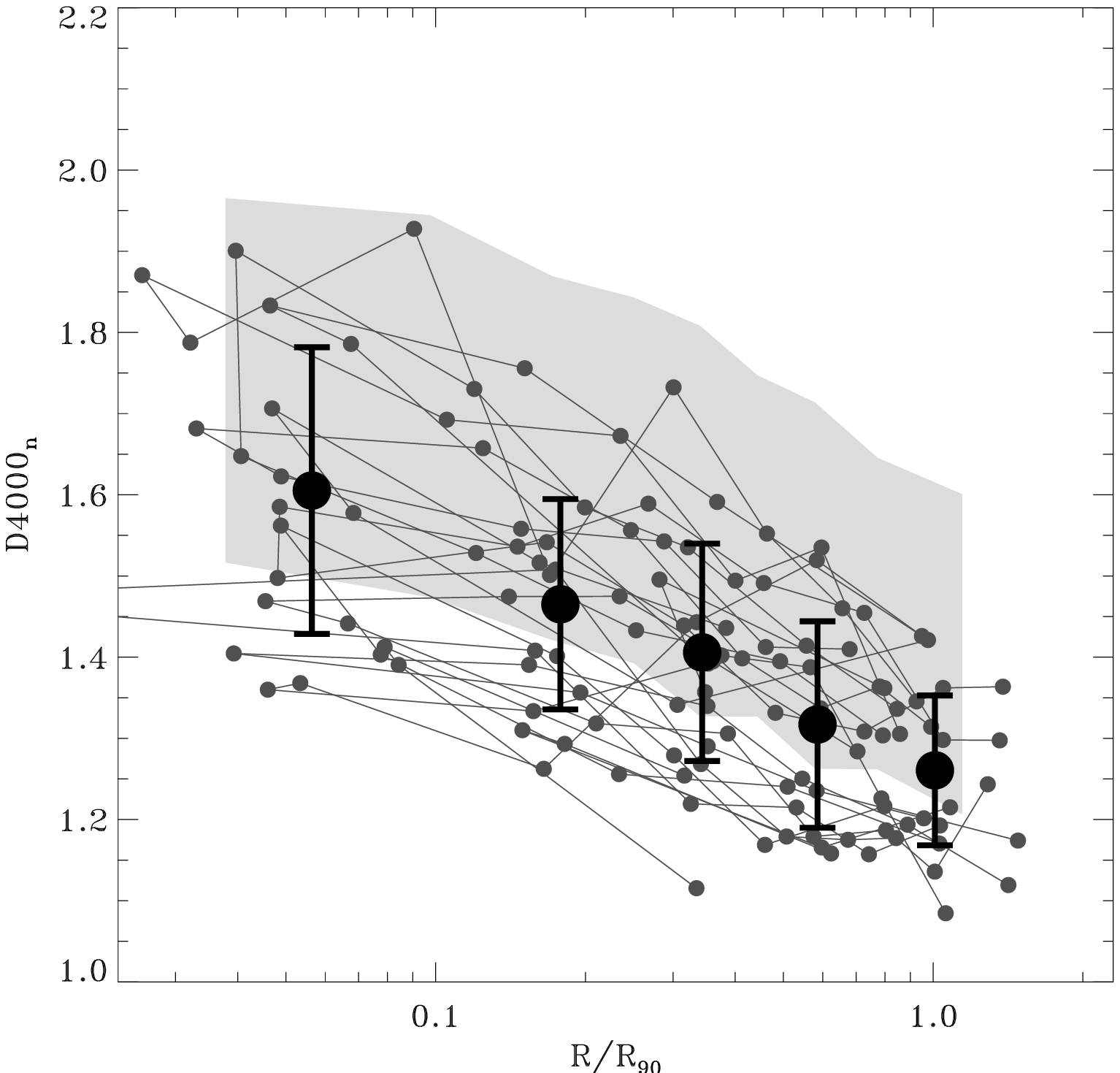}
\includegraphics[width=0.66\columnwidth]{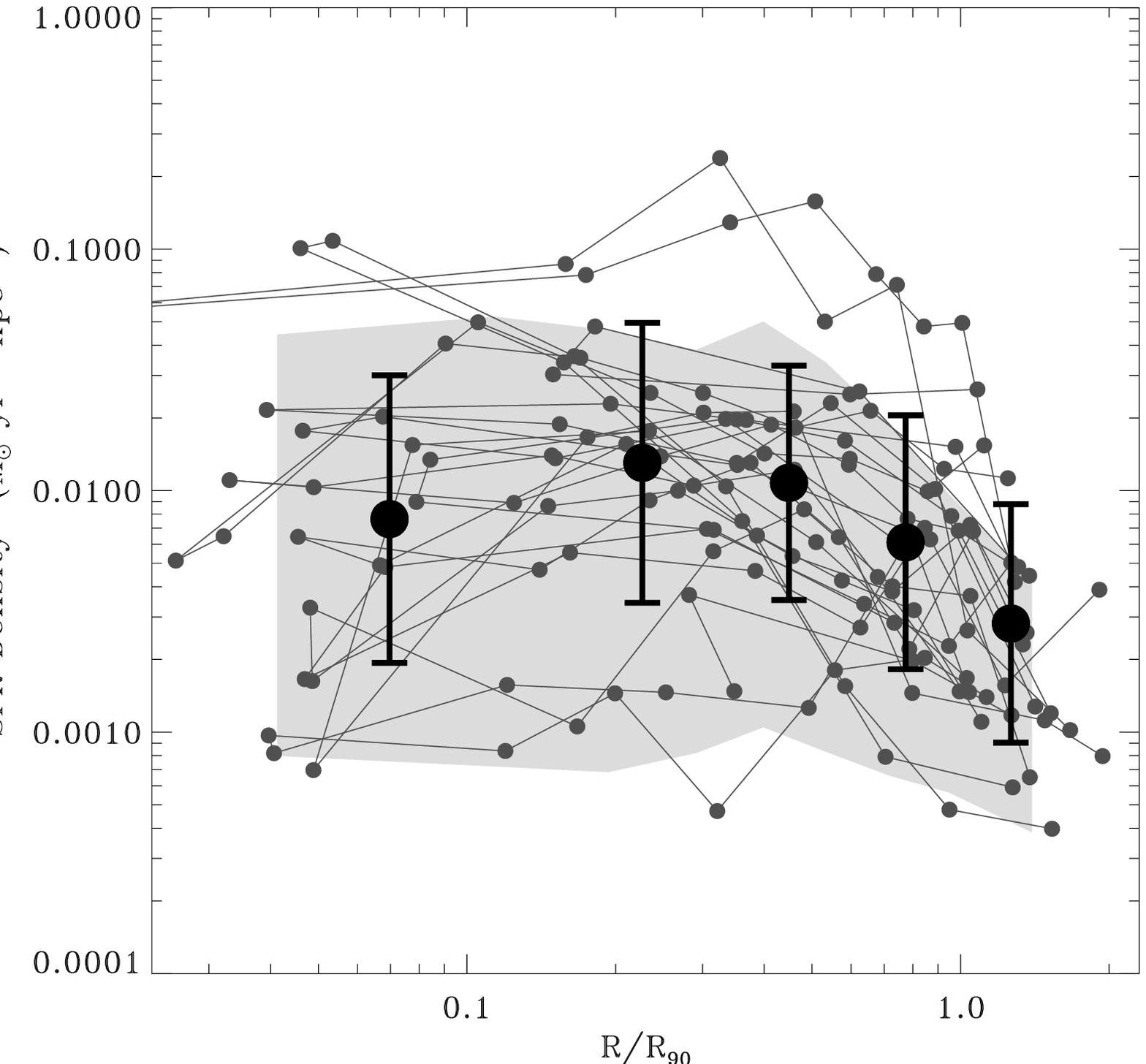}
\includegraphics[width=0.66\columnwidth]{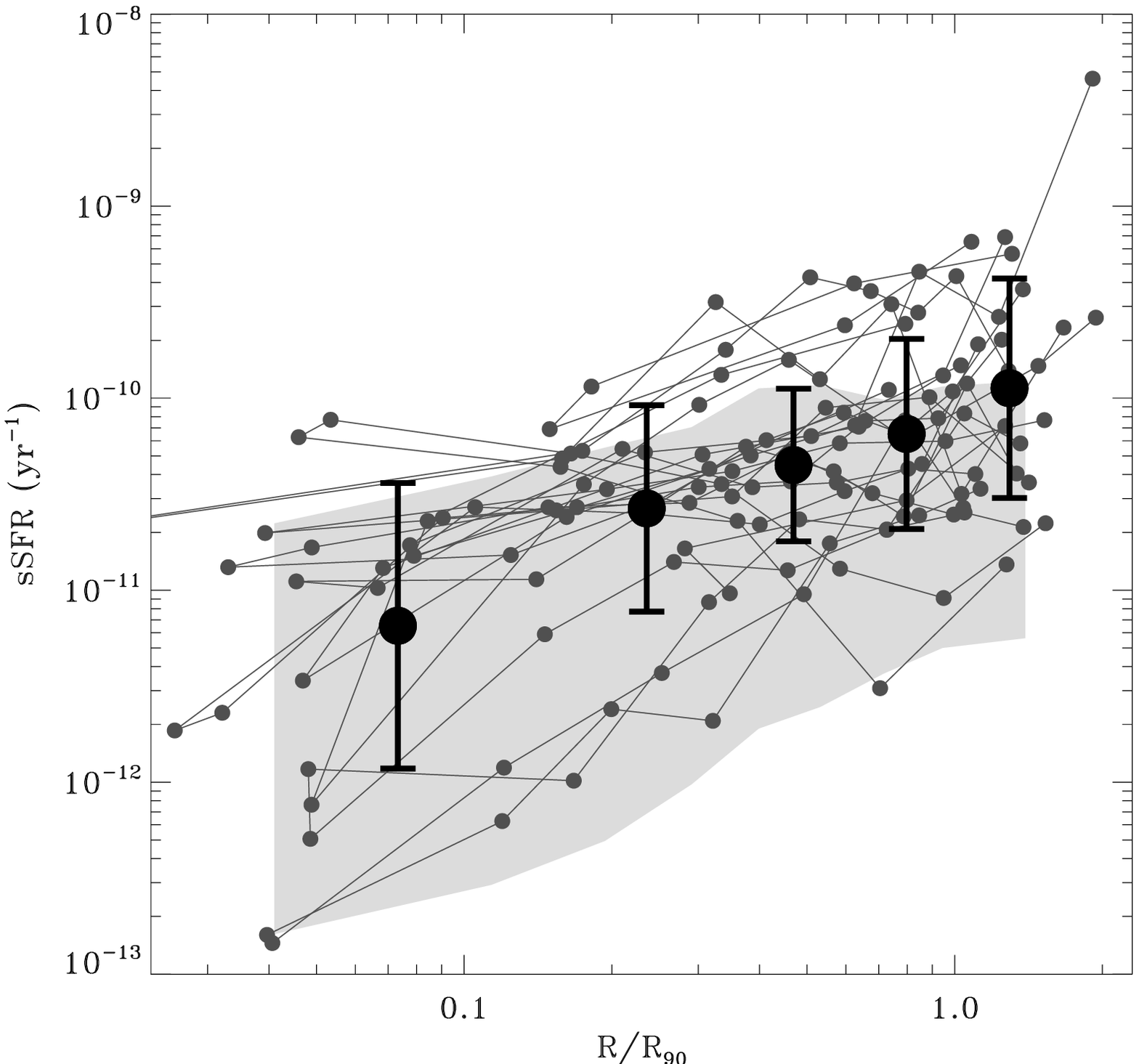}
\\
\includegraphics[width=0.66\columnwidth]{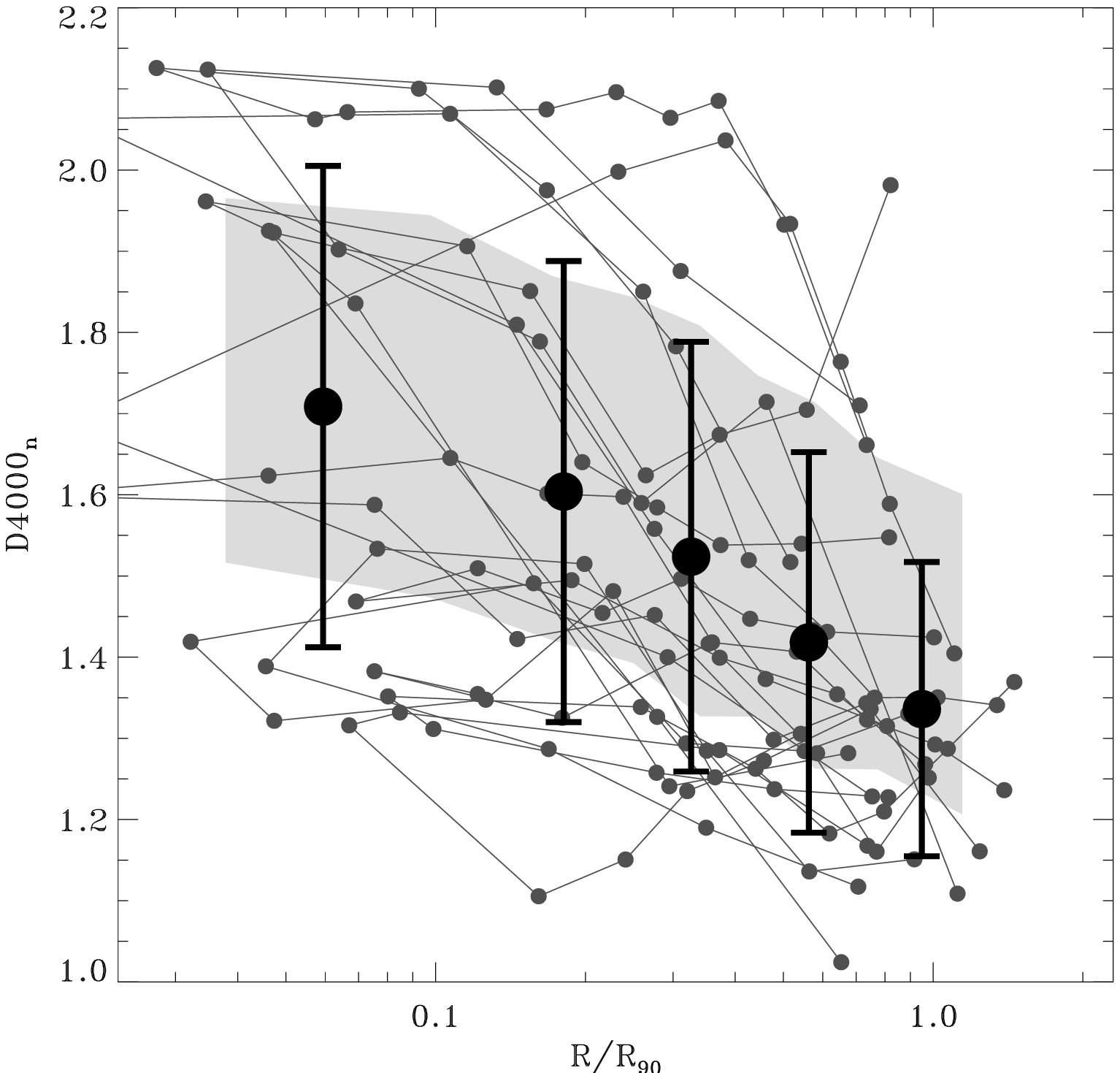}
\includegraphics[width=0.66\columnwidth]{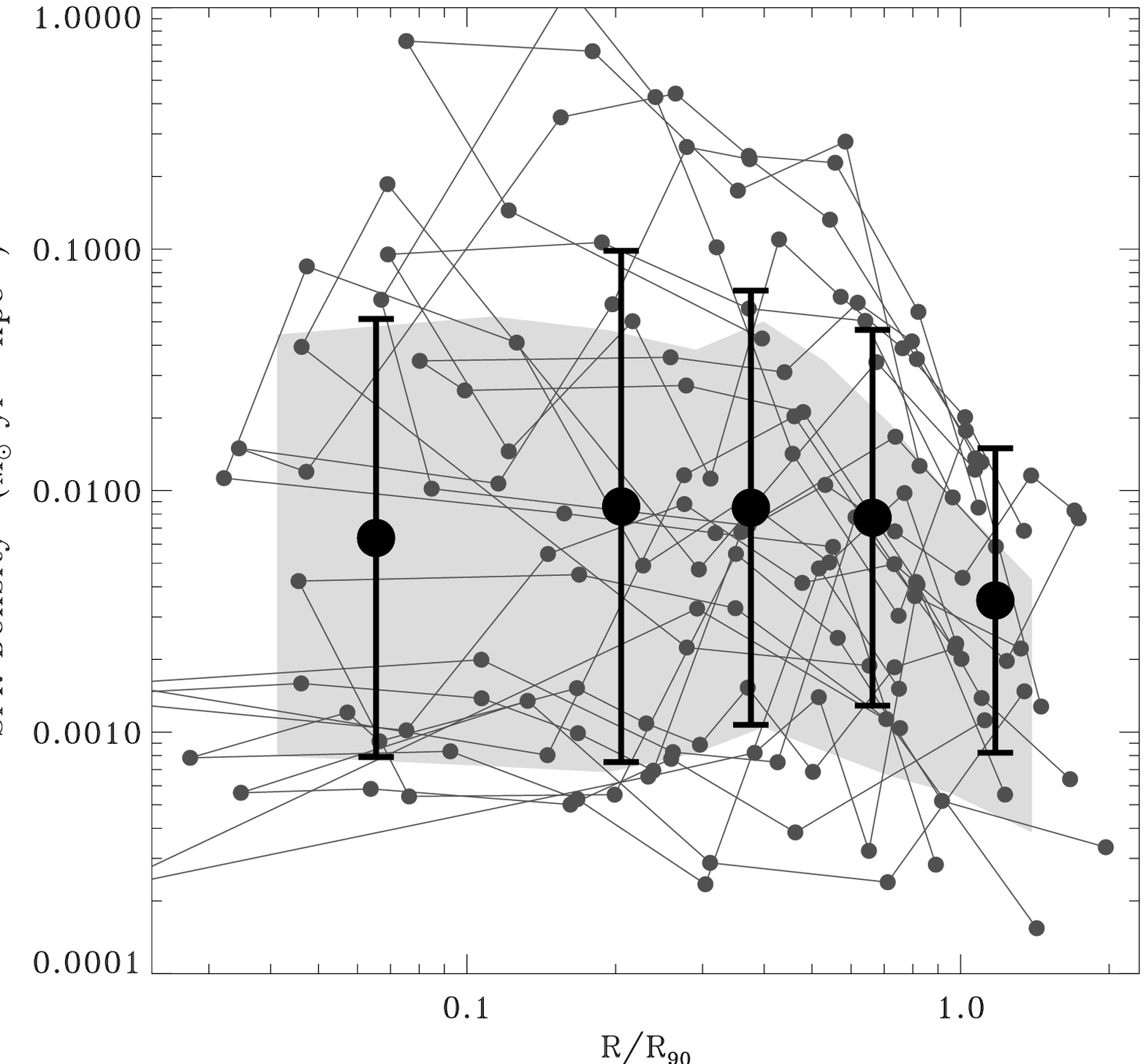}
\includegraphics[width=0.66\columnwidth]{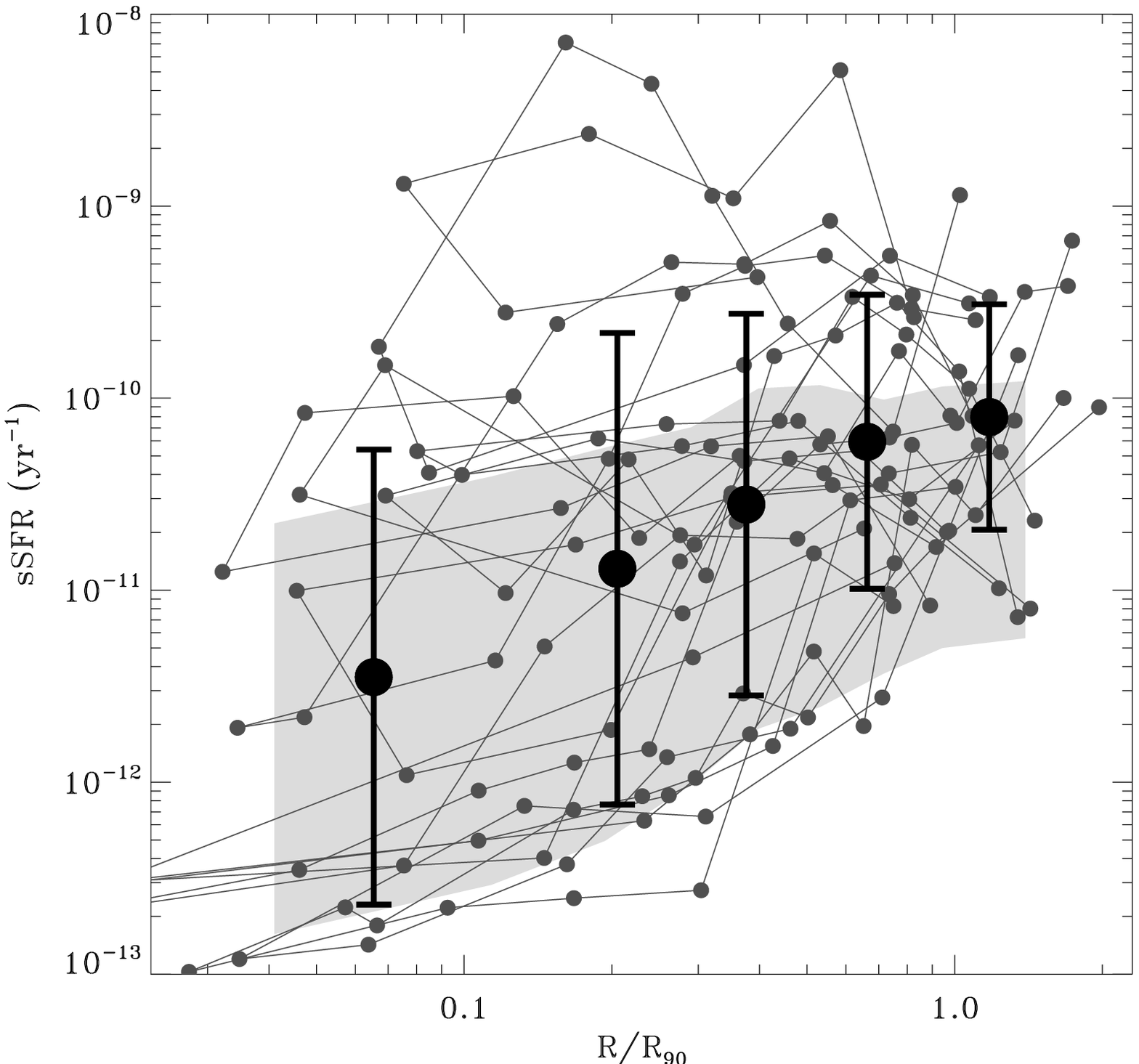}
\caption{\label{other_gradients} From left to right, $D4000_n$, SFR
  density, and sSFR gradients as a function of $R/R_{90}$, for galaxies
  in the large-metallicity-drop sample (top row),
  and those in the control sample (bottom row). Grey points show gradients
  with adjacent points connected by lines, as in Figure~10. Thick black
  points with error bars show the mean and 1-sigma values for bins of
  $\sim25$ points each. Light grey shading shows the $\pm1-\sigma$ region
occupied by the full GASS sample with extended star formation, divided 
  into radial bins of $\sim150$ points each, 
  and including only star-forming regions for $D4000_n$.
  We plot $D4000_n$ only for points with continuum S/N greater than 5 per angstrom, and
  uncertainty in $D4000_n$ less than 0.1. Typical uncertainties are less
  than 0.05.}
\end{figure*}

In Figure~\ref{other_gradients}, we plot, from left to right, the $D4000_n$ index, SFR
density (in $M_\odot$~yr$^{-1}$~kpc$^{-2}$), and sSFR (yr$^{-1}$) 
as a function of $R/R_{90}$,
for both the galaxies with large metallicity drops (top), and the
control sample (bottom). As in Figure~\ref{metal_gradients}, 
grey lines and dots connect the points for individual galaxies.
Black dots with error bars show the average and scatter in
bins of radius, for all the galaxies in each subsample. Light grey shading 
denotes the $\pm1\sigma$ region occupied by our {\it entire}
119-galaxy GASS sample with extended star formation,
as a function of radius. We note that $D4000_n$ averages for the full sample are calculated 
only from points with detected star-formation, to ensure the curves reflect the 
same set of galaxies (and individual points) as those in the other two panels.

The top panels of Figure~\ref{other_gradients} show that, for
the large-drop galaxies, the radial profiles of all three quantities 
show rather similar trends from galaxy to galaxy. 
$D4000_n$ generally decreases monotonically from center to outskirts (signifying
decreasing stellar population ages), specific
star formation rate {\it increases} monotonically towards the outside, 
and SFR density seems to exhibit largely flat radial profiles for most galaxies
(though the precise level of SFR density varies considerably).

Moreover, it is clear by comparing to the underlying shaded region marking the
full GASS sample, that, while the trends for decreasing age and increasing sSFR with radius 
are typical of all star-forming galaxies in this mass range, large-drop galaxies appear
overall younger and are building up stellar mass faster than the typical GASS 
galaxy. Large-drop galaxies have $D4000_n$ shifted an average of 0.16 lower 
than the full-sample mean, and sSFR is fully 0.5~dex higher, which translates to
mass-doubling times three times shorter---as low as 1~Gyr at the galaxy edges in the most 
extreme cases. Both of these
differences are statistically significant in all radial bins. SFR density is also 
higher by an average of 0.27~dex, though the difference in the innermost bin is not 
significant.

We note that all three plots for the large-drop galaxies seem remarkably
similar to the detailed properties of UGC8802 described in \citet{moran10}, 
where we argued that the combination of flat SFR density
profile and declining $D4000_n$ profile could easily be replicated by a
toy model featuring a recent episode of constant star formation spread
evenly across the galaxy, on top of an older stellar population that
built up most of the pre-existing stellar mass at an earlier time
($>1-2$ Gyr ago). The fact that $D4000_n$ indicates stellar populations that are
everywhere younger than the full-sample average, while sSFR is everywhere
higher than the full-sample average, lends support to this scenario.

One of the most striking features in Figure~\ref{other_gradients}
is the large difference in scatter between the large-drop and control samples, 
visible in all three measured quantities. In contrast to the
fairly uniform profiles of the large-drop galaxies, those of the control sample
are quite heterogeneous. For each of $D4000_n$, SFR density, and sSFR, the 
formal rms scatter at every radius bin in the large-drop sample is very 
nearly half what we measure in the corresponding control sample bin (with 
the precise ratio ranging from 50\% to 65\%). The level of heterogeneity 
displayed by the control sample is essentially indistinguishable from
that of the full GASS sample, and the two samples have the same mean values as
well, with one exception discussed below.
Thus, the spread in star formation rates and stellar population ages of 
the control sample can be considered typical for star-forming galaxies in this
mass range, while those for the large-drop sample are abnormally uniform and offset.

Only in the outermost bin (or two bins, for sSFR) do the mean properties of the
control sample of galaxies differ significantly from those of the full GASS sample.
In each measured parameter, the mean value of this outermost bin (or two) is instead 
consistent with that of the large-drop sample. In other words, the
high sSFRs and young ages at the very edges of our big-drop galaxies are
indistinguishable from those of other galaxies selected to have the same
global NUV-{\it r}, $M_*$, and $\mu_*$, {\it even though the metallicities are dramatically
different.} What this means is that we cannot definitively link the high sSFRs and young ages
to the high HI content and/or low metallicities at these locations, because we cannot exclude the
possibility that these features are generic for galaxies with this combination of
global properties. In contrast, the uniformly younger and more vigorously star-forming 
profiles at lower radii can perhaps {\it only} be attributed to the high HI and/or 
low outer metallicities, since
all the other parameters are identical between the two samples.

Can these two statements, seemingly at odds, be tied together into a uniform picture
of what is happening in the high HI, large-metallicity drop galaxies? We believe the
following scenario is plausible:

\bigskip

In galaxies with a large HI content, it has been shown that much of the gas often resides
beyond the optical disk of the galaxy \citep[e.g.,][]{bigiel08}. During the process of galaxy
growth, much of this gas must eventually be transported inward, where it will form
molecular gas and then stars. At the same time, it has been known for a long time
that radial flows of gas can naturally lead to metallicity gradients \citep[e.g.,][]{lacey85},
and that an `inside--out' buildup of galaxy disks---like we 
see in the sSFR profiles of big drop galaxies---may be required to explain the observed 
strengths of gradients \citep{boissier00}. 
We speculate that it is this transport of gas inward that sets
the level of metallicity suppression at the outer edges of our galaxies: star-forming
gas at the optical edge of the galaxy is diluted in simple proportion to the total 
amount of gas residing in the extended reservoir, and the continuous flow of such gas
serves to {\it keep} the metallicity low.

Under this scenario, the homogeneity of the radial profiles in the 
`large-drop' sample might arise because the star formation across the entire
galaxy becomes dominated by the dynamics of this inwardly-transported gas. 
A dense flow of gas, by providing ample fuel to all corners of the galaxy,
could well act to suppress the normal spatial variations in star formation rate seen in 
more typical galaxies, and at the same time cause a period of intense disk-building that
elevates sSFRs everywhere, including at the outskirts where the disk is building fastest.
In the control sample, small amounts of gas may also be building up
the disk outskirts, but with a
more-nearly-complete metal enrichment (or, rather, a less effective dilution of metals).
In these systems, however, the lower quantities of gas
involved are not enough to homogenize the star formation rates or
stellar population age gradients along the lines seen for the
large-drop objects. 

Let us return briefly to the question of flat inner metallicity gradients,
because it is puzzling why gradients should `saturate' and flatten out so effectively in
our high-mass galaxies. 
Generically, our observational result implies that the new metals produced by 
star formation are everywhere (except at the outskirts) precisely counterbalanced by 
either the net inflow of gas, or outflowing winds, or both. Since inner metallicity profiles
are flat for galaxies with both high and low $f_{HI}$ (Figure~\ref{metal_gradients}), 
this balance presumably holds over a range of gas densities. Such an equilibrium in
metallicity may seem implausible, but on the other hand could just be
the latest of many such observational `conspiracies' in the properties of galaxies.
In any case, it appears that models of the Milky Way with varying assumptions about cosmological 
gas infall and the details of `inside--out' formation can produce a range of flat to sloping 
inner gradients \citep[][and references therein]{colavitti09}. A full comparison to
models will not be undertaken here, but it is important to recognize that at least
some models {\it can} reproduce both the flat inner gradients and the steep outer drops we see.

To recap, we have clearly shown that galaxies with large outer metal drops not only
have high HI content, but show evidence that this gas is currently involved in a substantial
and widespread disk-building phase. Though the fates of these galaxies
as their gas depletes and metallicities rise is still unclear, it does seem to be the case
that a low outer metallicity is a sensitive signpost or thermometer for the presence of large
amounts of active gas in a galaxy. Given that 10\% of our overall GASS sample appears to be 
in this phase actively building disks, even though GASS probes a stellar mass range where such
activity is thought to be {\it decreasing}, it will be interesting
to see whether the abundance of these systems is compatible with theoretical models of 
galaxy formation and evolution.

\section{Summary}

In this paper, we have presented results on the gas-phase metallicities of 
galaxies in GASS, a homogeneous, representative sample spread evenly in stellar mass with
$M_*>10^{10}$M$_\odot$. We have described the following key results:

\begin{itemize}
\item We find strikingly flat metallicity profiles across massive galaxies out to typically the 
R$_{90}$ radius. The flat overall character of these profiles runs counter
to the prevailing view that galaxies typically have declining
metallicity gradients of varying slope \citep[e.g.,][]{oey93, zaritsky94, moustakas10}. 

\item However, our results are not inconsistent with these previous studies,
since we find evidence that inner metallicity profiles may vary systematically with galaxy mass.
Metallicity profiles which decline steadily with radius 
are observed only at the lowest masses in our sample (Log$(M_*)<10.2$), and an
inspection of other samples \citep{moustakas10,garnett97} suggests that they, too, find
declining gradients predominantly in lower mass galaxies.

\item Beyond $\sim R_{90}$, in many galaxies we observe a sharp downturn 
in metallicity. These occur in galaxies in all stellar mass bins, and the
largest drops of greater than $\sim0.25$~dex are observed in about 10\% of GASS galaxies.
Remarkably, we find that the magnitude of the outer metallicity drop is extremely 
well correlated with the {\it total} HI content of the galaxy.

\item We describe a correlation between local stellar mass 
density and metallicity, similar to the global mass-metallicity relation, 
valid across all galaxies in our sample. A very similar relation is found for
the centers of SDSS galaxies, which span a wider range in stellar mass, and obviously 
occupy regions with very different physical conditions. We speculate that local gas
mass fraction is strongly correlated with the degree to which metal enrichment has
occurred in all of these regions.

\item We examine in detail the subsample 
of galaxies exhibiting the most extreme outer drops in metallicity, and 
hence highest HI content. By examining
the radial profiles of their stellar population ages and star formation rate 
densities, we have shown that these galaxies are actively growing their 
stellar disks at a rate higher than the typical GASS galaxy, by a factor of about three
at all radii.

\end{itemize}

Thus, we argue that much of the recent stellar mass growth in these galaxies, as well
as the suppressed metallicities in their outskirts, can be directly linked to the 
inward transport of relatively pristine gas from beyond the galaxies' stellar disks. 
More specifically,
we speculate that the gas-phase metallicity at the outer edges of galaxy disks
acts as a thermometer of sorts for measuring the total gas reservoirs
of galaxies. Given recent work showing extended flat gradients in both mergers 
\citep{kewley10, rupke10} and
galaxies with extended HI disks \citep{werk11, bresolin09}, 
this relationship may not hold so
cleanly in dynamically disturbed systems. Likewise, it is unclear whether the local 
$\mu_*$-metallicity relation holds in such systems. With the detailed rotation curves we
measure from our longslit spectra, and our final, larger, sample of 300 galaxies,
we hope to address the role of galaxy dynamics in a future paper. 

The surprising result that outer metallicity is linked so tightly to
galaxy HI content may have been glimpsed before, by \citet{zaritsky94}. 
However, due to the much smaller sample of galaxies involved,
they were not able to determine if their observed correlation between
HI mass and mid-disk metallicity---not outer metallicity, which could be
another factor obscuring the trend---was fundamental, or due to
induced correlations through other galaxy properties like mass or
Hubble type. In a similar vein, \citet{skillman96} reported elevated metallicities
in HI-deficient Virgo-cluster spirals, but again the sample was small. 
In our much larger sample, we have been able to show that
the relation between HI fraction and outer metallicity is the tightest one, and that 
a model where outer metallicity is sensitive to the presence of an 
extended HI reservoir is consistent with all the data.

We have proposed a simple model where roughly 10\% of all massive disk galaxies today
sit in an extended HI reservoir that slowly drains inward, but testing this idea
will require comprehensive {\it resolved} information on
galaxies' HI reservoirs, as well as a better understanding of how galaxies are fed
by their surrounding large-scale structure and its associated inter-galactic gas. 
We are pursuing a number of extensions to GASS that may provide at least some
of this information, and we hope to revisit these questions soon.

\acknowledgements
S.M. wishes to thank R. Yates, M. Fall, E. Skillman, and D. Thilker for valuable discussions.
Observations reported here were obtained in part at the MMT Observatory, a facility
operated jointly by the Smithsonian Institution and the University of
Arizona. MMT telescope time was granted by the University of
Arizona, and by NOAO, through the Telescope 
System Instrumentation Program (TSIP). TSIP is funded by NSF.

The Arecibo Observatory is part of the National Astronomy and Ionosphere Center,
which is operated by Cornell University under a cooperative agreement
with the National Science Foundation. 
This work includes observations carried out with the IRAM 30-m telescope. IRAM
is supported by INSU/CNRS (France), MPG (Germany), and IGN (Spain).
{\it GALEX (Galaxy Evolution Explorer)} is a NASA Small Explorer, launched in
April 2003. We acknowledge NASA's support for construction, operation, and
science analysis for the {\it GALEX} mission.
Funding for the SDSS has been provided by the Alfred P.
Sloan Foundation, the Participating Institutions, the National Science
Foundation, the U.S. Department of Energy, the National Aeronautics
and Space Administration, the Japanese Monbukagakusho, the Max Planck
Society, and the Higher Education Funding Council for England.

\appendix
\section{Metallicity Profiles of Individual Galaxies}
In this Appendix, we provide plots showing the individual radial profiles of
metallicity, on the O3N2 scale, for all 100 of our objects that have quality measurements at eight 
or more discreet points. In the following pages, profiles are arranged in
order of ascending stellar mass, from top left to bottom right of each page.
The stellar mass of each galaxy is listed at the lower left of each plot, below
the internal GASS id number for each galaxy, given to aid future cross-referencing against
the GASS catalogs, when they are made available at http://www.mpa-garching.mpg.de/GASS/.
Individual points, as in earlier figures, are connected by lines, with the two
sides of each profile folded onto the same $R/R_{90}$ x-axis. We attempt to
place most galaxies on a common x-axis ranging to $1.8R/R_{90}$, but a few
objects with particularly extended data are plotted with an x-axis extending
to $2.3R/R_{90}$. Galaxies
that were included  in our large-drop sample in \S~5 are marked with an asterisk
at the top right of the plot; only ten of the thirteen galaxies are shown, as 
the other three had less than 8 measured points (but can still be seen in
Figure~\ref{metal_gradients}). The errorbars drawn for each point reflect
only statistical errors, to better illustrate variations in S/N; we remind
the reader that the typical systematic uncertainty is 0.07~dex, considerably 
larger than most of the errorbars shown. 
Upper limits are indicated by a downward facing arrow. Though they were 
included in Figures~\ref{metal_r90} and \ref{metal_by_mass}, for clarity we have excluded
from these plots any point where the limit on $12+Log(O/H)$ is higher than 
8.8, as very little information is added by such weak constraints.

\begin{figure}[t]
\includegraphics[width=\columnwidth]{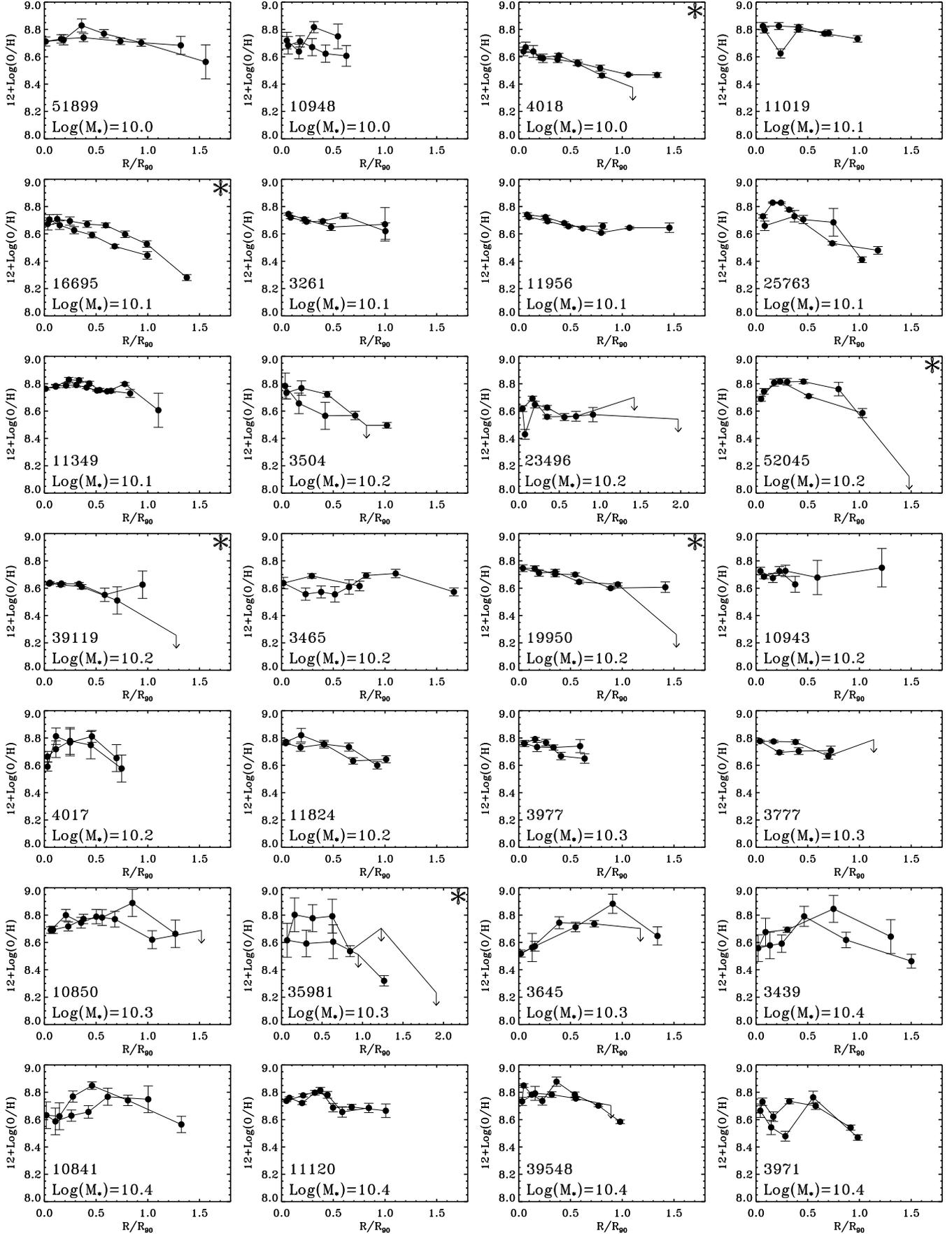}
\caption{Individual metallicity profiles for all 100 galaxies with
at least eight measured points, as described in the text.}
\end{figure}
\begin{figure}[t]
\includegraphics[width=\columnwidth]{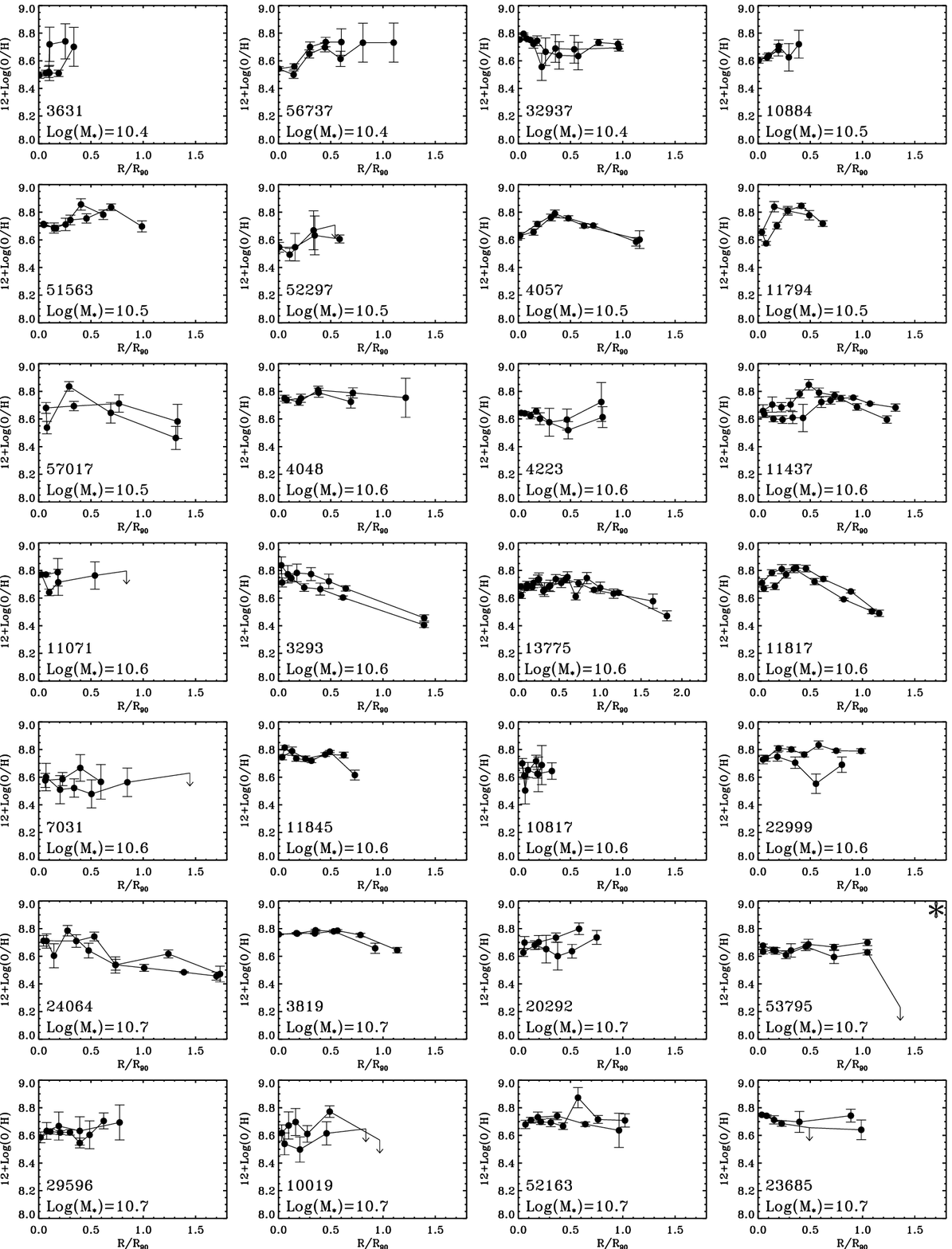}
\caption{Individual metallicity profiles for all 100 galaxies with
at least eight measured points, continued.}
\end{figure}
\begin{figure}[t]
\includegraphics[width=\columnwidth]{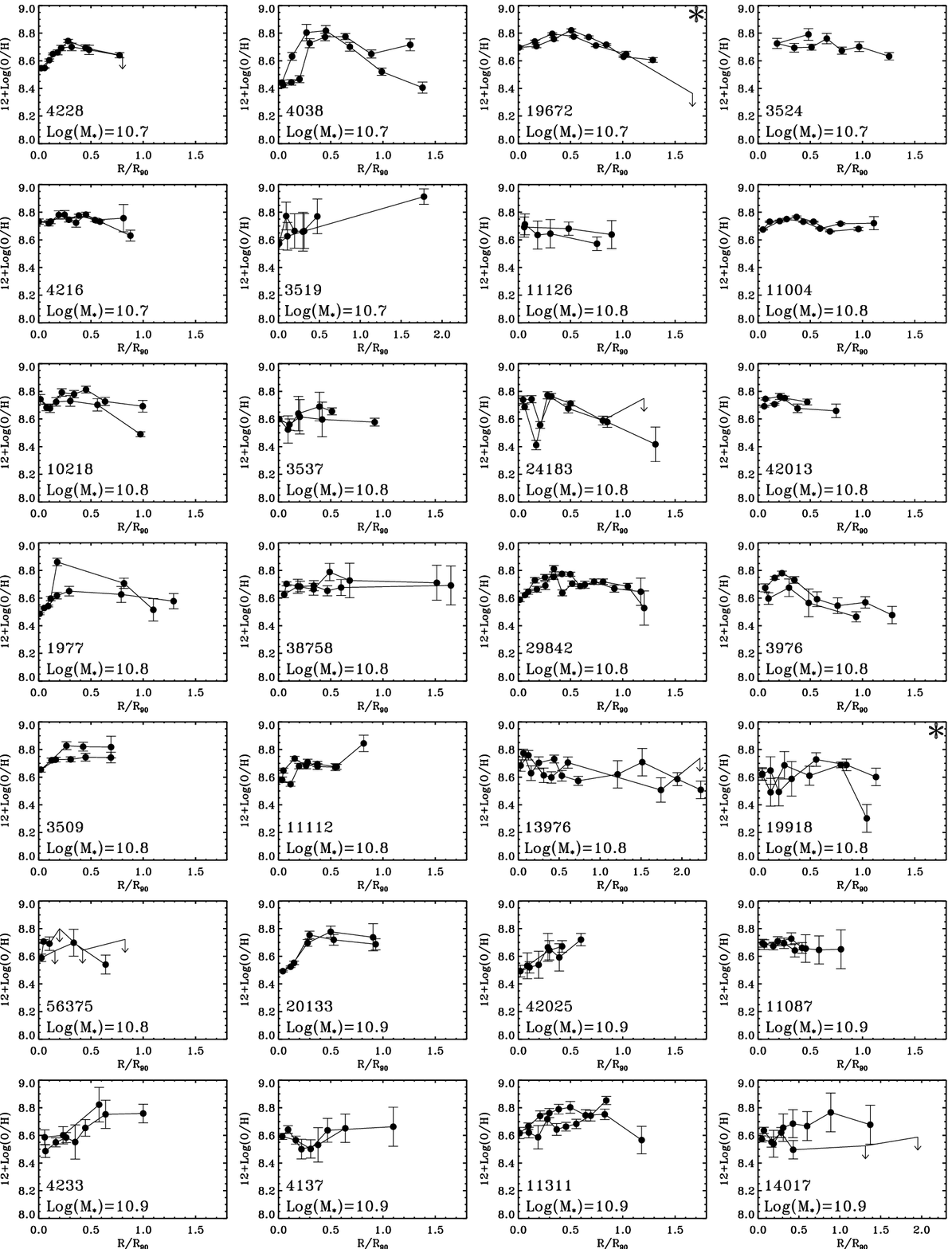}
\caption{Individual metallicity profiles for all 100 galaxies with
at least eight measured points, continued.}
\end{figure}
\begin{figure}[t]
\includegraphics[width=\columnwidth]{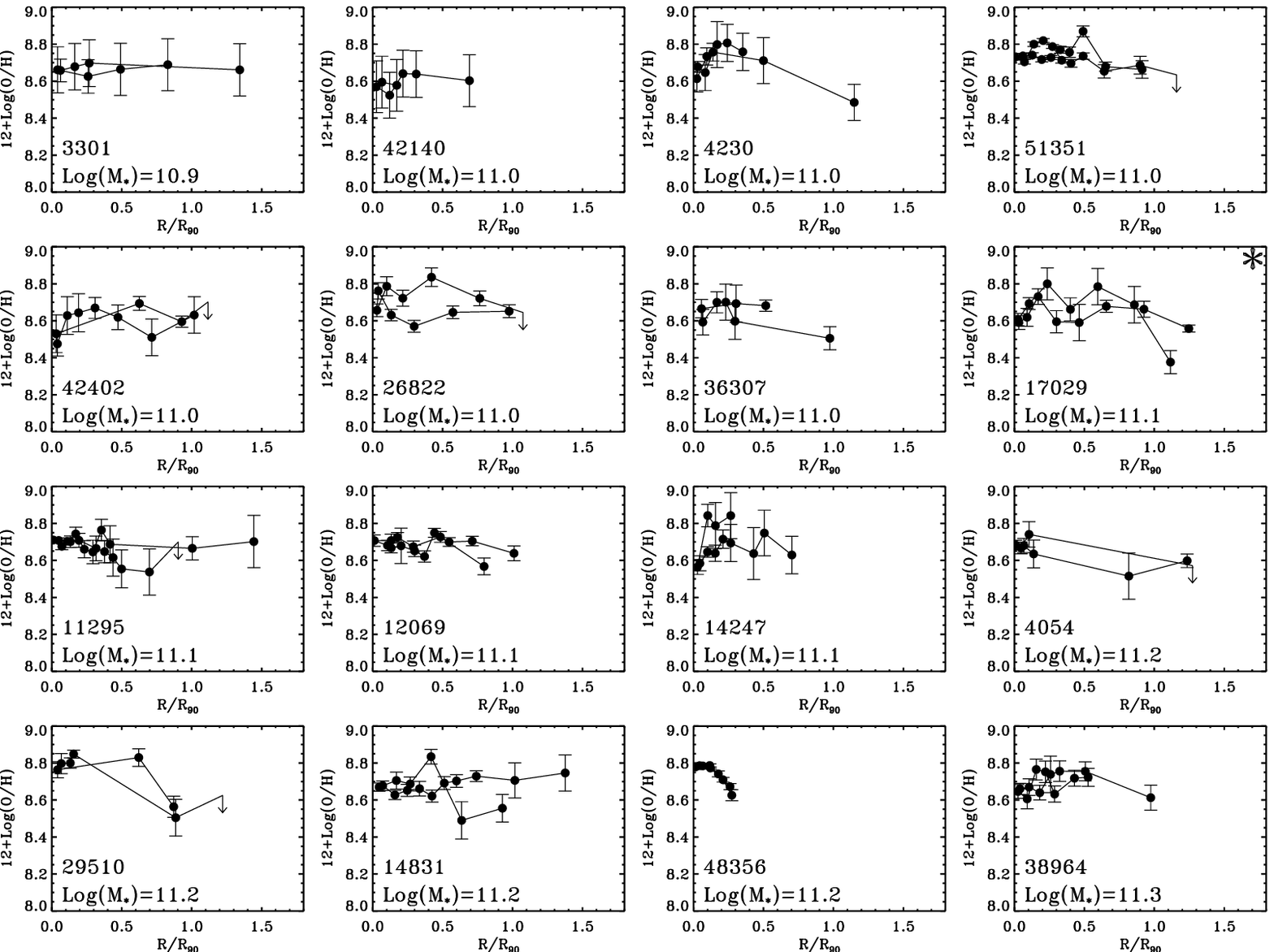}
\caption{Individual metallicity profiles for all 100 galaxies with
at least eight measured points, continued.}
\end{figure}

\end{document}